\documentclass[useAMS,usenatbib]{mn2e} %{aastex}
\usepackage{graphics,times,array,ctable,amsmath,amssymb}

\newcommand{\HI}{H{\sc ~i}}
\newcommand{\HII}{H{\sc ~ii}}
\newcommand{\HeI}{He{\sc ~i}}
\newcommand{\HeII}{He{\sc ~ii}}
\newcommand{\HeIII}{He{\sc ~iii}}

\title[Thermal Evolution of the IGM]{Models of the Thermal Evolution of the Intergalactic Medium After Reionization}
\author[Upton Sanderbeck, D'Aloisio, \& McQuinn]{Phoebe R. Upton Sanderbeck$^{1}$\thanks{Email: phoebe2@u.washington.edu}, Anson D'Aloisio$^1$, and Matthew J. McQuinn$^1$\\
$^1$Department of Astronomy, University of Washington}
\begin{document}
\maketitle

\begin{abstract}
Recent years have brought more precise temperature measurements of the low-density intergalactic medium (IGM).  These new measurements constrain the processes that heated the IGM, such as the reionization of \HI\ and of \HeII.  We present a semi-analytical model for the thermal history of the IGM that follows the photoheating history of primordial gas. Our model adopts a multi-zone approach that, compared to previous models, more accurately captures the inhomogeneous heating and cooling of the IGM during patchy reionization. We compare our model with recent temperature measurements spanning $z= 1.6-4.8$, finding that these measurements are consistent with scenarios in which the \HeII\ was reionized at $z= 3-4$ by quasars.  Significantly longer duration or higher redshift \HeII\ reionization scenarios are ruled out by the measurements.  For hydrogen reionization, we find that only low redshift and high temperature scenarios are excluded.  For example, a model in which the IGM was heated to $30,000\;$K when an ionization front passed, and with hydrogen reionization occurring over $6<z<9$, is ruled out. Finally, we place constraints on how much heating could owe to TeV blazars, cosmic rays, and other nonstandard mechanisms. We find that by $z= 2$ a maximum of $1$~eV of additional heat could be injected per baryon over standard photoheating-only models, with this limit becoming $\lesssim 0.5$ eV at $z>3$.
\end{abstract}


\section{Introduction}

When the ultraviolet photons from the first galaxies reionized the neutral
hydrogen in the Universe, a process that is thought to have occurred over $z\sim 6-12$ \citep{fan06, wmap, robertson10,planck15}, the intergalactic medium (IGM) was significantly heated from likely hundreds of degrees to $\sim 10^4~$K by
the excess energy above $1~$Rydberg from each photoionization. After the reionization of hydrogen and the first electron of helium, the next and last heating event of unshocked gas
is thought to have been the reionization of the second electron of helium, likely by quasars' ionizing emissions at $z\approx 3-4$ \citep{reimers97, theuns02, mcquinn09}.  After each reionization event, the evolution of the IGM temperature is driven by photoionization heating of residual bound electrons, the expansion of the Universe, and by well-characterized cooling processes of primordial gas \citep{1997MNRAS.292...27H, 2003ApJ...596....9H}.  However, it is possible that other processes contributed to the thermal evolution of cosmic gas, such as heating from cosmic rays \citep{samui05, lacki13}, from the intergalactic absorption of blazar TeV photons \citep[][]{chang12, puchwein12}, or from broadband intergalactic dust absorption \citep{inoue08}.

Hydrogen Ly$\alpha$ forest absorption in high-resolution, high signal-to-noise quasar spectra is the primary dataset used for measuring the temperature of the low-density IGM, having been used to place constraints over a broad range in redshift, $1.5<z<4.8$ (recently by \citealt{lidz10, becker, boera}). Temperature measurements from the Ly$\alpha$ forest are based on the Ly$\alpha$ absorption features of a colder IGM having sharper features than those of a warmer one; the spatial distribution of higher temperature gas is more extended because of higher pressures, and the absorption of hotter gas is additionally smoothed by enhanced thermal broadening.  Many methods to estimate the temperature that exploit these effects have been devised, including directly fitting for the widths of Ly$\alpha$ forest absorption lines as a function of \HI\ column (e.g., \citealt{2000MNRAS.318..817S, ricotti00, mcdonald01, bolton14}), measuring the exponential suppression that owes to temperature at high wavenumbers in the Ly$\alpha$ forest power spectrum \citep{zaldarriaga01}, and convolving Ly$\alpha$ forest spectra with wavelet (and wavelet-like) filters that return larger values for sharper features \citep{lidz10, becker,theunszaroubi,theuns02b,zaldar02}.  Regardless of the temperature estimation method, analyses must mock observe Ly$\alpha$ forest spectra synthesized from cosmological hydrodynamic simulations to calibrate their temperature estimates. 

There was a significant focus on measuring the IGM temperature from the Ly$\alpha$ forest around the the turn of the century \citep{2000MNRAS.318..817S, ricotti00, mcdonald01}.  The main conclusion from these earlier measurements was that the temperature at $z\sim 3$ was too high for hydrogen reionization (and a contemporaneous full reionization of helium) to be the most recent heating event and likely implied either (1) \HeII\ reionization had occurred around $z\sim3$ \citep{2003ApJ...596....9H} or (2) some non-standard heating process was operating \citep{samui05}. The measurements of \citet{2000MNRAS.318..817S} showed evidence for an increase and then decrease in temperature straddling $z\sim3$, for which the simplest explanation is that it owes to \HeII\ reionization \citep{theuns02}. However, this thermal bump was not obviously present in other contemporaneous measurements \citep{ricotti00, mcdonald01, zaldarriaga01}.  In the last several years, there has been a renewed focus on temperature estimates.  Of particular note, \citet{becker} measured the thermal history over $2<z<4.8$ at a specific redshift-dependent density chosen to minimize errors, allowing for a higher precision temperature estimate.  This measurement qualitatively agrees with the earlier result of \citet{2000MNRAS.318..817S} that showed evidence for \HeII\ reionization.  The \citet{becker} measurement has been complimented by a subsequent effort by \citet{boera} extending temperature measurements down to $z=1.5$.\footnote{ In addition, \citet{bolton12} measured the temperature in the proximity region of seven quasars at $5.8<z<6.4$.  The special location of this measurement, likely a region where the hydrogen was reionized early on in the reionization process \citep{lidz07} and where the \HeII\ was reionized by the quasar make this measurement more difficult to interpret, but see \citet{bolton12}.} 
  The \citet{becker} measurement has also been confirmed over $2<z<3.2$ using other techniques \citep{garzilli12, bolton14}. Furthermore, in addition to the measurements of temperature at a single density, the scaling of temperature with density has recently been constrained well at $z=2.4$ \citep{2012ApJ...757L..30R, bolton14}, adding to previous less-certain estimates over a range of redshifts \citep{ricotti00, mcdonald01, 2000MNRAS.318..817S}. Motivated by this recent spate of IGM temperature measurements, this study presents the most realistic models to date of the thermal history of the IGM in the standard scenario in which photoionization is the dominant energy source.  A byproduct of these studies is bounds on other heating and cooling processes

This paper is organized as follows. Section~\ref{sec:tempmeasurements} discusses the temperature measurements that we use to compare our models. Section~\ref{sec:model} describes a model for the temperature evolution of the low-density intergalactic medium after hydrogen reionization. Section~\ref{sec:beforeHeIIreion} compares this model with the highest redshift  temperature measurements at $z\sim4.5$ to constrain hydrogen reionization.  Section~\ref{sec:general} then discusses the thermal evolution at lower redshifts when the heating from \HeII\ reionization is likely to be important.  We place constraints on non-standard heating processes in Section~\ref{sec:nonstandard}. All calculations assume a flat $\Lambda$CDM Universe with $h = 0.7$,
$\Omega_m = 0.3$, $\Omega_b = 0.045$, and $Y_{\rm He} = 0.25$.

\section{Temperature Measurements}
\label{sec:tempmeasurements}

\begin{figure}
\begin{center}
\resizebox{9.0cm}{!}{\includegraphics{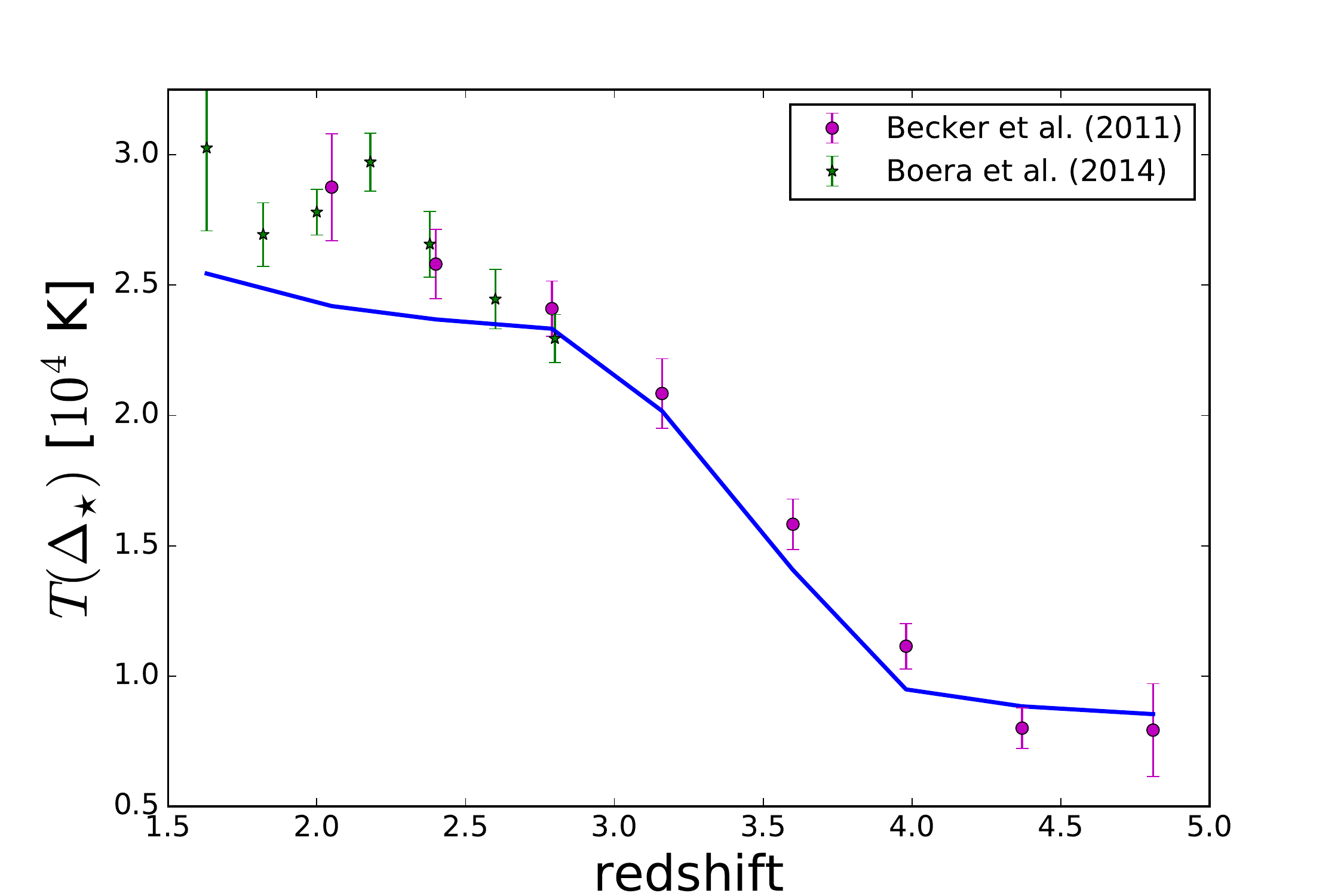}}\\
\end{center}
\caption{The IGM temperature at the critical densities, $\Delta_{\star}$, of \citet{becker} and \citet{boera}. The solid curve corresponds to our model with \HeII\ reionization spanning $2.8<z_{\rm rei}<4$ (see \S~\ref{sec:general}). The points with error bars are the measurements of ~\citet{becker} and ~\citet{boera}. The corresponding densities of the Becker measurements are $\Delta_{\star} = { 5.69,4.39,3.35,2.62,2.02,1.64,1.40,1.23}$ at $z = {2.05,2.40,2.79,3.16,3.60,3.98,4.37,4.81}$. The densities of the Boera measurements are $\Delta_{\star} = {5.13,4.55,4.11,3.74,3.39,3.08,2.84}$ at $z={1.63,1.82,2.00,2.18,2.38,2.60,2.80}$.}
\label{T_of_Delta}
\end{figure}

We primarily compare our models with the temperature measurements of \citet{becker}, spanning $2.1\leq z_{\rm em}\leq4.8$ and of \citet{boera}, spanning $1.6\leq z_{\rm em}\leq2.8$. Like other Ly$\alpha$ forest temperature measurements, these studies used that the width of Ly$\alpha$ absorption features is sensitive to thermal and pressure broadening, with sharper absorption features arising from lower temperature gas.  Specifically, \citet{becker} and \citet{boera} adopted the ``curvature statistic,'' defined as:
\begin{equation}
\kappa\equiv\frac{F''}{\left[1+(F')^2\right]^{3/2}},
\label{eqn:kappa}
\end{equation}
where $F'$ and $F''$ represent the first and second derivatives of the flux with respect to the velocity along a sightline. Colder regions have higher $\langle |\kappa | \rangle$ than hotter ones. Like with the closely-related wavelet method \citep{theunszaroubi, lidz10} as well as the power spectrum method \citep{zaldarriaga01}, the curvature method does not require fitting for individual absorption lines. 

The \citet{becker} and \citet{boera} temperature measurements, featured in Figure~\ref{T_of_Delta}, quote smaller errors than in previous IGM temperature studies.  The increased precision arises from the realization in \citet{becker} of a nearly one-to-one relationship between the mean curvature, $\langle |\kappa| \rangle$, and the temperature at a critical density of the Ly$\alpha$ forest, $\Delta_*$ (see Fig.~6 of \citealt{becker}).  (Throughout, we denote densities in units of the mean baryonic density as $\Delta$.)  This one-to-one relationship allows for scaling the measured curvature, $\left<|\kappa|\right>$, into a measured $T(\Delta_{\star})$.  The physical reason for this correspondence is that the Ly$\alpha$ forest lines with optical depth $\tau_{\rm Ly\alpha} \sim 1$ are most sensitive to the IGM temperature, and the critical density that produces $\tau_{\rm Ly\alpha} \sim 1$ depends on redshift.  Previous studies used other methods to constrain the temperature at $\Delta = 1$, $T_0$, and its power-law index, $\gamma-1$, using the parameterization
\begin{equation}
T(\Delta)=T_0\;\Delta^{\gamma-1}.
\end{equation}
This approach resulted in larger errors because the sensitivities of these methods are often greatest at densities significantly different from $\Delta=1$.  However, for this reason, the \citet{becker} measurement does not directly constrain $T_0$, so a model for $\gamma -1$ is needed in order to extrapolate from $T(\Delta_{\star})$ to $T_0$.

\begin{figure}
\begin{center}
\hspace{-1.2cm}
\resizebox{9.5cm}{!}{\includegraphics{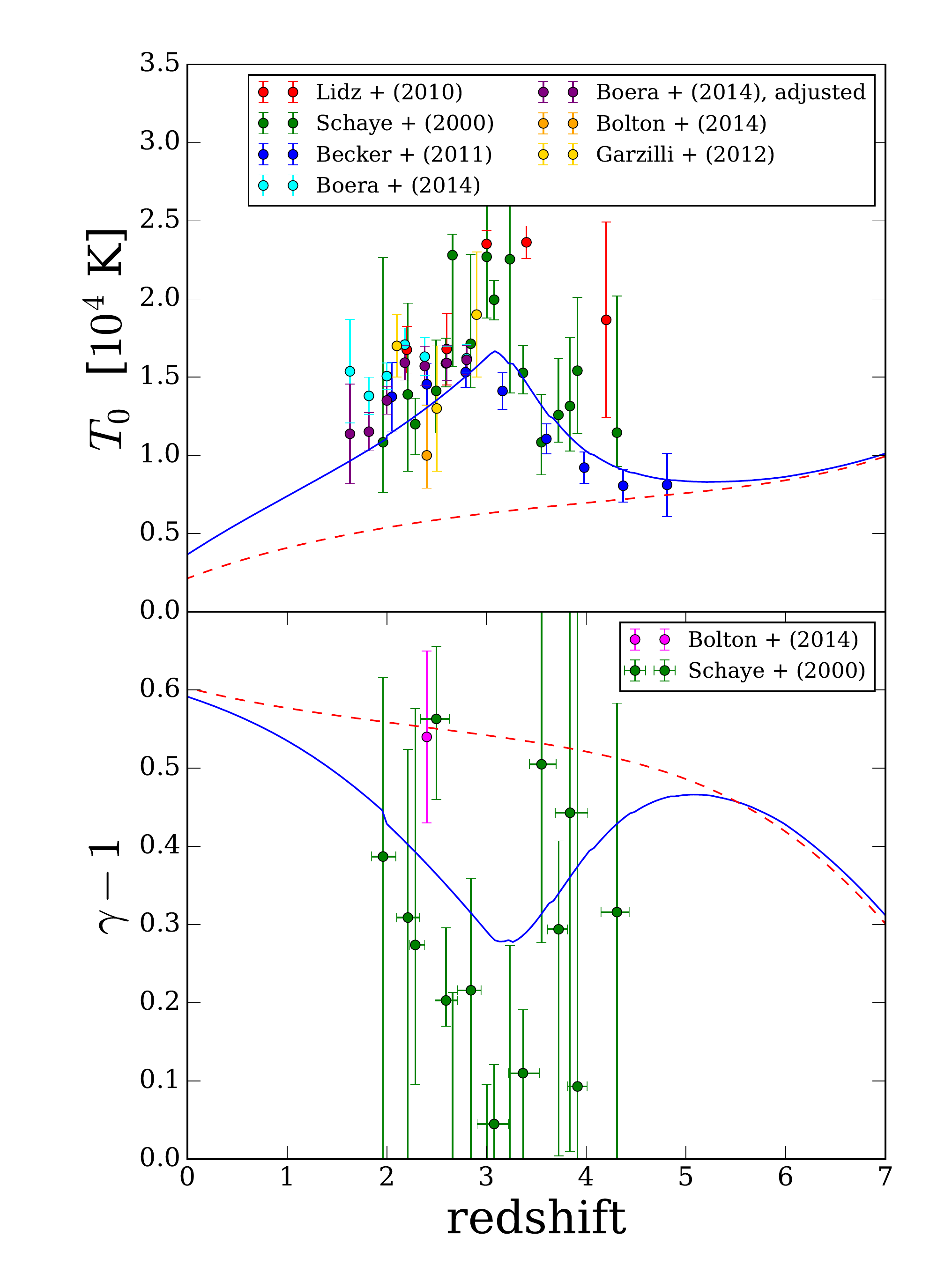}}\\
\end{center}
\caption{Measurements of $T_0$ (top panel) and $\gamma-1$ (bottom panel) from \citet{schaye00}, \citet[fixing $\gamma-1$ to that in our model]{lidz10}, \citet{garzilli12}, and \citet{bolton14}.  Also shown in the top panel is the \citet{becker} and \citet{boera} estimates using the $\gamma-1$ of the blue model curve to extrapolate from their $\widehat{T}(\Delta_*)$ to $T_0$. The blue model curve is our fiducial heating history described in \S~\ref{sec:general} with $\alpha_{\rm QSO}=1.7$ and $\alpha_{\rm bkgd} = 1.0$, whereas the red curve assumes the same parameters but sets the $>4~$Ry background to zero such that the \HeII\ is never reionized.
\label{othertemps}}
\end{figure}

Lastly, we note that the \citet{becker} measurements agree with other attempts using different methodologies. At $z\sim 2.5$, the \citet{becker} measurements are consistent with the results of \citet{garzilli12}, which used wavelet plus the flux PDF methods, and \citet{bolton14}, which used line-fitting methods.  The \citet{becker} measurements fall at just slightly lower temperatures than the earlier study of \citet{schaye00}, but show similar trends.   However, they appear to be discrepant with the measurements reported in \citet{lidz10}.  Figure~\ref{othertemps} shows a compilation of $T_0$ and $\gamma-1$ measurements, alongside our fiducial model with \HeII\ reionization (blue curve) and the same model but where the $>4~$Ry background is set to zero such that the \HeII\ is never reionized (red curve).  Both models are described in the ensuing sections.  To make the comparison fair, for the \citet{becker}, \citet{boera}, and \citet{lidz10} data points in the top panel we take the $\gamma-1$ values from our fiducial model.  The actual measurements of $T(\Delta_*)$ from which the former two studies extrapolate, along with this theoretical model, are shown in Figure~\ref{T_of_Delta}.\footnote{\citet{lidz10} also published estimates of $T_0$ that marginalize over $\gamma-1$, resulting in larger error bars.}

\section{Modeling the temperature of the IGM}
\label{sec:model}

\subsection{The processes that affect the temperature}
\label{ss:methodology}

The temperature evolution of a Lagrangian fluid element is described by the following equation \citep{1997MNRAS.292...27H}
\begin{equation}
\frac{dT}{dt}=-2HT+\frac{2T}{3\,\Delta}\frac{d
\Delta}{dt}-\frac{T}{n_{\rm tot}}\frac{dn_{\rm tot}}{dt}+\frac{2}{3k_Bn_{\rm tot}}\frac{dQ}{dt},
\label{eqn:dTdz}
\end{equation}
where $H$ is the Hubble parameter and $n_{\rm tot}$ is the total number density of free ``baryonic'' particles (electrons and ions).  Equation~(\ref{eqn:dTdz}) does not include the shock heating from structure formation that heats some low density gas. We will comment on this omission later.

The first term on the right hand side of equation~(\ref{eqn:dTdz}) accounts for adiabatic cooling due to the expansion of the Universe.  At densities near the cosmic mean -- densities where the temperature is measured -- this expansion term tends to be the dominant coolant at $z<6$, whereas at higher redshifts Compton cooling off of CMB photons becomes important for ionized gas. 

The second term on the right hand side of equation~(\ref{eqn:dTdz})  describes the adiabatic evolution due to collapsing overdensities and expanding voids. We model the density evolution of a gas element as a Zeldovich pancake \citep{zeldovich70} such that
\begin{equation}
\Delta(a) = \left[1- \lambda G(a) \right]^{-1},
\end{equation}
where $G(a)$ is the growth factor (which over most redshifts considered scales as $a$) and $\lambda$ is a parameter that is tuned to achieve a particular density at a given redshift.  While much of the moderately overdense IGM that concerns us (i.e. $1 \leq \Delta \leq \Delta_*$) is well modeled by this restrictive form for $\Delta(a)$, we note that the actual thermal evolution is weakly sensitive to the history of $\Delta(a)$, as shown explicitly in \citet{2015arXiv150507875M}.  We find that our results do not change if we assume different functional forms for $\Delta(a)$, such as that in spherical collapse (the approach in \citealt{2009ApJ...701...94F}).

The third term on the right hand side of equation~(\ref{eqn:dTdz}) is relatively unimportant, accounting for the change in the number of particles in the thermal bath. This term only affects our calculations during \HeII\ reionization.  We implement a 4\% drop in the temperature when the \HeII\ becomes ionized to account for this term.

The fourth term on the right hand side of equation~(\ref{eqn:dTdz}) is where all the details of our model lie, encoding all other heating and cooling processes. In standard models in which photoheating is the only heating process, the fourth term can be written as
\begin{equation}
\frac{dQ}{dt} = \overbrace{\sum_X \frac{dQ_{{\rm photo}, X}}{dt}}^{\rm heating} + \overbrace{\frac{dQ_{\rm Compton}}{dt} + \sum_i\sum_X R_{i,X} n_{e} n_{X} }^{\rm cooling},
\end{equation}
where $dQ_{{\rm photo}, X}/dt$ is the photoheating rate of ion $X$, $dQ_{\rm Compton}/dt$ is the Compton cooling rate, and $R_{i,X}$ is the cooling rate coefficient for ion $X$ and cooling mechanism $i$, including recombination cooling, free-free cooling, and collisional cooling.  For the cooling terms, our calculations adopt the rate coefficients in the appendix of \citet{1997MNRAS.292...27H} as well as the free-free cooling rate of ~\citet{rybickilightman}.  \citet{2015arXiv150507875M} discusses the relative importance of these processes.  For the photoheating term, more modeling is required.

\subsubsection{Modeling photoheating}

There are two regimes for the photoheating rate, $dQ_{\rm photo}/dt$: (1) The photoheating when an ion is being reionized, and (2) the optically thin photoheating that an ion experiences thereafter. The latter post-reionization rate is given by
\begin{eqnarray}
\frac{dQ_{{\rm photo}, X}}{dt} &=& n_{X}\int_{\nu_{X}}^{\infty}\frac{d\nu}{h\nu}4\pi J_{\nu}\sigma_{X}(\nu) \times [h\nu-h\nu_{X}], \label{eqn:photoheating}\\
&\approx&\frac{h\,\nu_{X}}{\gamma_X - 1 + \alpha}\alpha_{A,X}n_{\tilde{X}}n_{e},
\label{eqn:photoheating2}
\end{eqnarray}
where $\alpha_{A,X}$ is the CASE A recombination coefficient associated with the transition $X{\rm I} \rightarrow X$, $X \in$ \{ \HI\ , \HeI\ , \HeII \} and $\tilde{X}\in$ \{\rm H, He\}, $\nu_X$ is the frequency associated with the ionizing potential of species $X$, $\sigma_X(\nu)$ is the photoionization cross section, and $\gamma_X$ is the approximate power-law index of $\sigma_X(\nu)$ for which we take values of $\{2.8, \, 1.7, \,2.8\}$ for \{\HI, \HeI, \HeII \}.  Equation~(\ref{eqn:photoheating2}) results from the approximation of photoionization equilibrium with an ionizing background that has a power-law specific intensity of the form $J_{\nu}\propto \nu^{-\alpha}$, where $J_{\nu}$ is the average specific intensity [erg s$^{-1}$ Hz$^{-1}$ sr$^{-1}$ cm$^{-2}$].\footnote{ Equation~(\ref{eqn:photoheating2})  also ignores photoionizations into ionization state $X$.  This approximation is very accurate.  For example, take the case of $X = $\HeII. Once the \HeI\ is reionized, $x_{\rm HeI} \approx 10^{-5} x_{\rm HeII}$ assuming photoionization rates relevant for \HeI.  Thus, \HeII\  -- which exists in much higher fractions ($x_{\rm HeII}\sim10^{-2}$) -- is formed primarily by recombinations from HeIII and not by photoionizations of \HeI, as equation~(\ref{eqn:photoheating2}) assumes. }

The heating during reionization depends on when reionization processes occurred. It is thought that the reionization of \HI\ and \HeI\ (what we will refer to as ``hydrogen reionization'') was driven by POPII stars. Observations of the Ly$\alpha$ forest suggest that this process ended by $z\approx 6$ \citep{fan06, mcgreer15}, with measurements of the cosmic microwave background suggesting a mean redshift for this process of $z=8.8^{+1.7}_{-1.4}$ \citep{planck15}. \HeII\ is thought to have been reionized by quasars, with observations of the \HeII\ Lyman-$\alpha$ forest suggesting that this process completed around $z\approx 2.8$ \citep{mcquinnGP, worseck11, shull10}.\footnote{POPII stars do not produce many photons that can doubly ionize the \HeII.  Even if there existed a non-nuclear galactic source that did produce these photons, it is unlikely that they would escape the local interstellar medium.}  In the standard picture, the duration of the \HeII\ reionization process depended on the emissivity of quasars, which has been well measured $z\lesssim 3$ and the constraints are improving at higher redshifts \citep{hopkins07, willot10, 2012A&A...537A..16F, mcgreer13}.  Models of \HeII\ reionization by quasars predict that \HeII\ reionization should span a significant time interval, at least $z=3-4$ \citep{furlanetto07, mcquinn09, compostella13}, although recent measurements of quasar abundances suggest that an earlier timeline may be possible \citep{giallongo, madau15}.  We constrain the duration of \HeII\ reionization here using its effect on the temperature.

Hydrogen and helium reionization heat the IGM with a higher rate than the optically thin photoheating rate \citep[eqn.~\ref{eqn:photoheating2}; e.g.][]{1999ApJ...520L..13A, trac08, mcquinn09}.  The amount of heating from these processes has been quantified by previous numerical studies. \citet{2012MNRAS.426.1349M} showed that the temperature after an HI ionization front passes is unlikely to be larger than $30,000~$K,  because of collisional cooling inside the ionization front, nor smaller than $20,000~$K, confirming the analytic arguments in \citet{miralda94} that it should fall around $20,000~$K.  (The exact temperature depends on the spectrum of the sources and speed of the ionization fronts.)
  We use these bounds in our models of hydrogen reionization discussed in Section~\ref{sec:beforeHeIIreion}.  In contrast, during \HeII\ reionization the heating is not localized just to the ionization front.  Instead, much of the heating occurs from a nearly uniform hard, $\gtrsim 200~$eV, background that has a long mean free path to be absorbed \citep{mcquinn09}.   Another difference with \HI\ reionization is that collisional cooling is not efficient in the \HeII\ ionization fronts such that the ionizing spectrum alone determines the amount of photoheating during \HeII\ reionization \citep{mcquinn09}.  We describe a physically motivated model for the photoheating during \HeII\ reionization in Section~\ref{sec:general}.

\subsection{Comparing our models with the temperature measurements}
\label{ss:comparing}

This section describes in detail how we compare our model predictions
with the \citet{becker} and \citet{boera} measurements. We warn the
reader in advance that this section is the most technical in the paper.
Readers not interested in how we infer the ``observed'' temperature
from our models can skip to the next section.

The problem this section addresses is that our models predict a
distribution of temperatures at every redshift and density rather than a
single temperature. Thus, we must understand how to map our models to the effective temperature \citet{becker} and \citet{boera} measured. In addition, even for histories with a single temperature at a given $\Delta$ (as in our models that
assume an instantaneous reionization), the measured temperature would
differ from the actual temperature because the \citet{becker} and
\citet{boera} measurement technique is calibrated for a specific
thermal history.\footnote{The pressure smoothing of the gas affects the
measured curvature, and pressure smoothing is sensitive to the
temperature over the previous dynamical time.  Indeed the curvature
statistic is more affected by pressure broadening than other Ly$\alpha$
forest temperature diagnostics \citep{ewaldetal}.}   Handling these
effects in full rigor would require running simulations that mimic the
inhomogeneous photoheating rates in our models, generating mock Ly
$\alpha$ forest skewers from the simulations, applying the curvature
statistic to them, and comparing directly to the \citet{becker} and
\citet{boera} curvature measurement.  Fortunately, running a simulation
for every model that we consider is unnecessary; we show here that the
\citet{becker} and \citet{boera} $T(\Delta_\star)$ estimates essentially
measure the average $T(\Delta_\star)$ of our models.

\begin{figure}
\begin{center}
\resizebox{9.5cm}{!}{\includegraphics{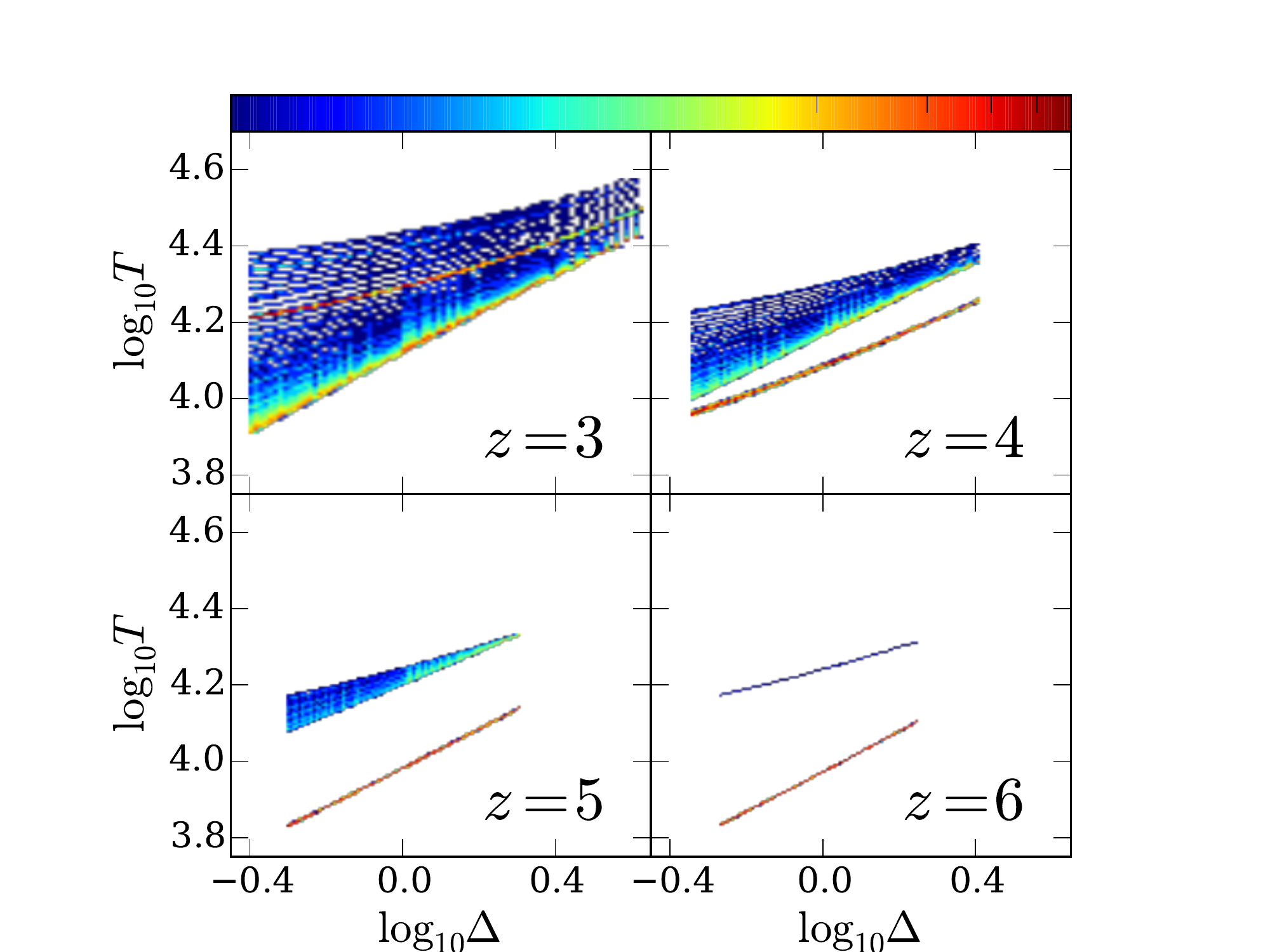}}
\end{center}
\caption{Predicted temperature distribution in a fiducial quasar
reionization model described in \S~\ref{sec:general} that uses the
\citet{hopkins07} emissivity history and that takes \HI\ reionization to
be instantaneous at $z=9$.  The colorbar denotes the logarithmic number of zones
that fall in the pixel such that red areas have approximately two orders
of magnitude times more points than dark blue.  Initially almost all
points fall on a single $T-\Delta$ relation.   Gradually as \HeII\
reionization proceeds a distribution of temperatures develops.  This
development is similar to that seen in radiative transfer simulations of \HeII\
reionization \citep{mcquinn09}.  Section~\ref{ss:comparing} motivates
why a simple average of $T(\Delta)$ yields the effective temperature
that has been measured. }
\label{Hires_Tarrs}
\end{figure}

Let us first focus on the effective temperature that the curvature
statistic would measure in our models, ignoring pressure smoothing for
the time being.  Generally the spatial temperature fluctuations in our models
have $\delta T/T<1$, with the size of fluctuations being smaller with
increasing $\Delta$.  Figure~\ref{Hires_Tarrs} shows the temperature-density distribution in the fiducial model described in \S~\ref{sec:general}.  The dispersion in temperature at fixed density owes to different regions being reionized at different times.   Even though the dispersion looks quite large, the fractional standard deviation at $\Delta_*$ in this model are only $\{0.14, 0.13, 0.14, 0.25\}$ at $z=\{3,\,4,\,5,\, 6\}$. The smallness of the standard deviation
means that any weighted average of the temperature would differ from the
actual mean temperature by a relatively small factor of $\sim T
\langle(\delta T/T)^2 \rangle$.  Indeed, as an example we find that if
we take the mean $T(\Delta)$ or the exponentiated mean of $\log
T(\Delta)$ in our fiducial model at $z=3$, the two values agree with astonishing accuracy, to $0.7$\%.
Thus, the effective temperature that Ly$\alpha$ forest temperature measurements
infer likely can be approximated by the average temperature.

To be more quantitative, we took the relation between $\langle |\kappa| \rangle$ and
$T(\Delta_{\star})$ measured from a suite of simulations in
\citet{becker}.  We then calculated
\begin{equation}
\overline{\langle |\kappa| \rangle} = \int dT_{\Delta_*} P(T_{\Delta_*})~
\langle |\kappa| \rangle (T_{\Delta_*}),
\end{equation}
 where $P(T_{\Delta_*})$ is the probability distribution of $T_{\Delta_*} \equiv T(\Delta_{\star})$ and $\langle |\kappa| \rangle (T_{\Delta_*})$ is the mean curvature in a region with temperature $T_{\Delta_*}$.
 Thus, $\overline{\langle |\kappa| \rangle}$ is the temperature-averaged curvature
the \citet{becker} method would infer from our models.\footnote{This assumes that the temperature fluctuations
we are modeling are coherent over much larger scales than the widths of
lines in the forest.}  Then, we used the \citet{becker} relation between
$\langle |\kappa| \rangle$ and $T(\Delta_{\star})$ to project $\overline{\langle |\kappa| \rangle}$
back to $T(\Delta_{\star})$.  Remarkably, we find that this agrees with
the average of $T(\Delta_{\star})$ to better than $1\%$ at $z\sim
3$ (when the temperature fluctuations are largest) in our fiducial model.  Thus, the curvature measurements constrain the average of $T(\Delta_{\star})$ in our
models.

Now let us focus on how to correct the temperature measurements to
account for the pressure smoothing (also known as ``Jeans smoothing'') in our models. The biases from not
having the correct pressure smoothing are most significant when the IGM
is heated over a time interval much less than the free-fall timescale ($\ll H^{-1} \Delta_{\star}^{-1/2}$), the timescale for a sound wave to travel the Jeans' scale (the relaxation time), such
as from a short \HeII\ reionization. The effect of pressure smoothing on $\langle |\kappa |\rangle$ was investigated in
\citet{becker}.  \citet{becker} calibrated the relation between
$T(\Delta_{\star})-\langle |\kappa| \rangle$ using a suite of simulations that had
relatively constant $T(\Delta_{\star})$ with time, much different than
the thermal histories in our models.  To investigate the potential bias from these rather artificial temperature histories,
\citet{becker} ran two simulations where global heating was injected in
a manner that emulates the heating from \HeII\ reionization (one where
\HeII\ reionization spanned $\Delta z \approx 0.5$ and another with $\Delta z
\approx 1$).  Then, \citet{becker} compared the estimated temperature
from applying the $\langle |\kappa| \rangle$ estimate and then their
$T(\Delta_{\star})-\langle |\kappa| \rangle$ relation to the actual temperature in the
simulation.  They found that this resulted in a moderate bias, that when
projected to $\Delta=0$ was $ \Delta T < 3000~$K, peaking near the end
of \HeII\ reionization, with a smaller bias at other redshifts (see Appendix~A).

  Because our models have a similar duration to the \HeII\
reionization simulatations in \citet[Appendix~A compares these
histories]{becker}, we correct the temperature values measured in
\citet{becker}, $\widehat {T}(\Delta_{\star})$. We multiply $\widehat {T}(\Delta_{\star})$ by the amount of bias due to ``Jeans smoothing" that they measure, $\Delta T$, by the ratio of the
temperature in our fiducial model, $T_{\rm fid}$, to their T15slow simulation with \HeII\ reionization, $T_{\rm
T15slow}$ (because the
\citealt{becker} simulations are at a bit higher temperatures).  Namely, we correct the temperature measurements using 
\begin{equation}
\widehat {T}(\Delta_{\star})_{\rm corrected} = \widehat
{T}(\Delta_{\star}) + \Delta T(z) \frac{T_{\rm fid}(\Delta_{\star})}{T_{\rm T15slow}(\Delta_{\star})} .
\end{equation}
We use the T15slow \HeII\ reionization simulation of \citet{becker} because it falls closest to our reionization
histories (as shown in Appendix A).  If we had instead used T15fast (which is shorter than in all of our models) the $\widehat {T}(\Delta_{\star})_{\rm corrected}$ increase by a maximum of $\approx 2500$ K, with this bias peaking at $z= 3.2$.  In addition, because the \citet{boera}
measurement focuses on $z< 2.8$, after \HeII\
reionization in our models, the pressure smoothing correction is
smaller:  $\approx1500$ K at $z=2.8$ and decreasing to
$<100$ K by $z=2.0$.

Since we find that comparing the average temperature is accurate at the percent level and our pressure smoothing corrections are at most a few thousand Kelvin, our approximate way of correcting for
these effects should be accurate to much better than the size of these corrections, more than sufficient for our purposes.  The smallness of these corrections also justifies ignoring the coupling of these two effects. Furthermore, $T_0$ -- the temperature at mean density -- is also a more intuitive quantity than
$T(\Delta_{\star})$.  To calculate $T_0$, we use the $\gamma$ in our models to
extrapolate both the measurements and the model to $\Delta=0$.  We fit for $\gamma$ using the mean temperature
and density at each redshift. This maintains the fractional difference
between the measurements and model, and we find that in all of our plots
it matters little which model we assume (and so we generally take the
fiducial set of parameters).  The model would matter if we considered models with and
without a late \HeII\ reionization in the same plot, but this is never
the case.

\section{Temperature before \HeII\ reionization}
\label{sec:beforeHeIIreion}

Our model of the temperature of the IGM begins at the epoch of hydrogen reionization when the IGM was likely heated to $20,000-30,000$ K as an ionization front passed. Before \HeII\ was reionized and in the simplistic limit of an instantaneous \HI\ (and \HeI) reionization, the uncertainties in the temperature can be parametrized by three numbers: the spectral index of the post-reionization ionizing background ($\alpha_{\rm bk}$), the final temperature of the IGM after hydrogen reionization ($T_{\rm rei}$), and the redshift of instantaneous hydrogen reionization ($z_{\rm rei}$).  In this section, we first explore these dependencies and then turn to a more realistic model for an extended and patchy hydrogen reionization.

\begin{figure}
\begin{center}
\hspace{-0.9cm}
\resizebox{9.2cm}{!}{\includegraphics{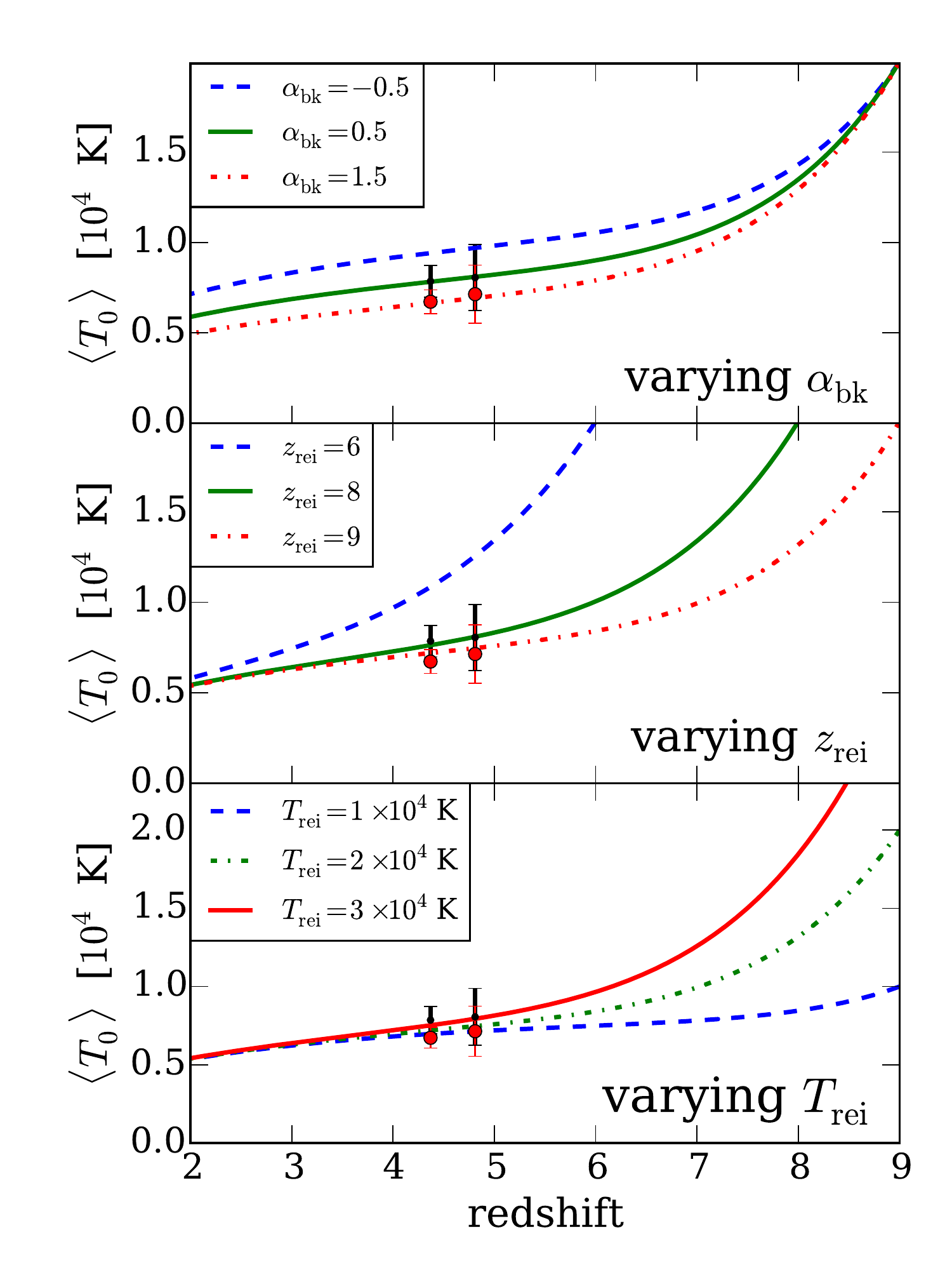}}\\
\end{center}
\caption{Average temperature of the IGM at the cosmic mean density in models that omit the photoheating from \HeII\ reionization and use $\alpha_{\rm bk}= 1$, $T_{\rm rei}=20,000$~K, and $z_{\rm rei} = 9$, unless stated otherwise. {\it Top panel:} Models that vary the spectral index of the ionizing background over the range of $-0.5<\alpha_{\rm bk}<1.5$. {\it Middle panel:} Models that vary the redshift of instantaneous \HI\  reionization.  {\it Bottom panel:} Models that vary the final temperature after hydrogen reionization.  The points with error bars are the two highest redshift temperature estimates of~\citet{becker}. The red points have been corrected for pressure smoothing and the black points are uncorrected.}  
\label{HI}
\end{figure}

The top panel in Figure~\ref{HI} shows the effect on the temperature history of the spectral index of the post-reionization ionizing background, $\alpha_{\rm bk}$, which affects the equilibrium photoheating rate after reionization.  Ignoring ionizing recombination radiation, this spectral index is set by the intrinsic spectral index of the sources, $\alpha_s$, and the logarithmic slope of the column-density distribution of intergalactic hydrogen absorbers, $\beta$, via the formula $\alpha_{\rm bk} \approx \alpha_s+3(\beta - 1)$ (valid at $z\gtrsim 3$ when $\lambda_{\rm MFP, HI}\ll cH^{-1}$, where $\lambda_{\rm MFP, HI}$ is the physical mean free path of $1~$Ry photons). The spectral index, $\alpha_s$, is measured to be approximately $1.5\pm 0.2$ if quasars dominate the ionizing background \citep{telfer02, shull12, lusso15}, and between 0 and 1.5 at $\sim 1~$Ry if it is stars \citep{faucher09, starburst99, conroy, bruzual03}. \citet{haardtmadau} shows a stellar population synthesis spectrum (Figure 15) where an effective slope of approximately $\alpha_s = 0.5$ appears most compatible with the population synthesis models over 1-4 Ry. In addition, $\alpha_s = 1$ was assumed in the UV background model of \citet{faucher09} based on population synthesis models.  However, this slope is likely hardened by things like stellar binarity \citep{stanway16} and the inevitably frequency-dependent escape fraction of ionizing photons. The value of $\beta$ is measured to be approximately $1.3\pm 0.2$ \citep{2010ApJ...721.1448S}, although it varies over the range of interest from steeper to flatter values \citep{prochaska10,2002ApJ...568L..71Z}.  In addition, direct recombination to the ground state can soften the effective spectral index that appears in the photoheating rate by $\approx 1$ unit \citep{faucher09}.  Motivated by these values, we consider the range of $-0.5 \leq \alpha_{\rm bk} \leq 1.5$. The top panel in Figure~\ref{HI} shows that $T(z)$ varies only by as much as $\Delta T \approx 2000~$K when we vary $\alpha_{\rm bk}$ over this large range.  The insensitivity of $T_0$ to $\alpha_{\rm bk}$ arises because the energy per optically thin photoionization is approximately proportional to $(2+\alpha_{\rm bk})^{-1}$, so that $ -0.5 \leq \alpha_{\rm bk} \leq 1.5$ results in a factor of $2.3$ variation in the photoheating rate, and at a maximum the temperature depends on the photoheating rate to the $0.6$ power  \citep{2015arXiv150507875M}.

The middle panel in Figure~\ref{HI} investigates how the timing of hydrogen reionization affects $T_0(z)$. The three thin curves assume, simplistically, that hydrogen reionization occurred instantaneously at $z_{\rm rei}=6,8,$ and $9$, heating the IGM to $2\times10^4~$K.  Due to Compton cooling, the temperature of these curves decreases quickly after reionization, erasing all memory of when reionization occurred by $z\sim 4$ for the cases with $z_{\rm rei} \geq 8$.  Only the $z_{\rm rei}=6$ curve results in a significantly different prediction for $T_0$ at $z=4.8$ than the highest-redshift \citealt{becker} measurement.     (We note that $\Delta_* \approx 1.2$ for the $z=4.8$ \citet{becker} estimate and so there is little extrapolation necessary to $T_0$.) 

The bottom panel in Figure~\ref{HI} shows temperature histories with different initial temperatures after \HI\ reionization, $T_{\rm rei}$. The IGM temperatures reached during reionization depend on the spectra of the sources as well as the speeds of the ionization fronts.  These temperatures are likely bracketed by $18,000$ and $25,000$ K \citep{miralda94, 2012MNRAS.426.1349M}.   One-dimensional radiative transfer calculations yield minimum temperatures of $\approx 18,000$ K for models in which the sources are much softer than is expected from stellar population synthesis models \citep{2012MNRAS.426.1349M}.  On the other hand, radiative cooling within the ionization front becomes very efficient at temperatures $\gtrsim 25,000$ K, making it difficult to achieve temperatures much higher than this.  However, for illustrative purposes we explore a broader range of temperatures in Figure~\ref{HI}, $T_{\rm rei}=10,000- 30,000$ K.  While unrealistic, the $10,000~$K case is similar to the temperature achieved in simulations that use optically thin photoheating rates.  The curves show the case of an instantaneous reionization at $z_{\rm rei} = 9$.

\begin{figure}
\begin{center}
\hspace{-0.9cm}
\resizebox{9.2cm}{!}{\includegraphics{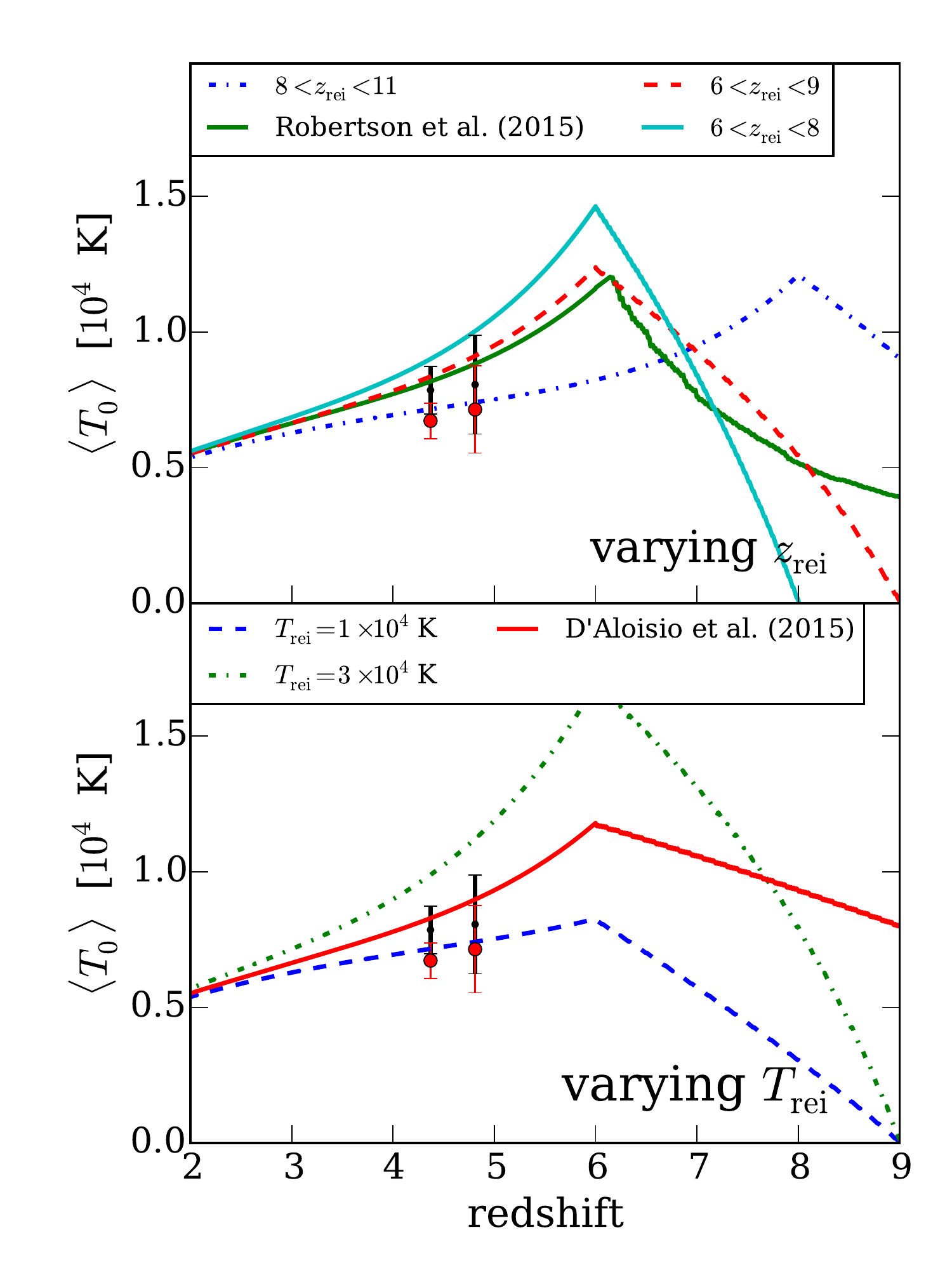}}\\
\end{center}
\caption{Average temperature of the IGM at the cosmic mean density in our models that use $\alpha_{\rm bk}= 1$ and omit the photoheating from \HeII\ reionization. {\it Top panel:} Models that vary the redshifts and durations of \HI\ reionization with $T_{\rm rei}= 20,000~$K.  The durations correspond to $6<z_{\rm rei}<8$, $6<z_{\rm rei}<9$, $8<z_{\rm rei}<11$, as well as the synthesis ionization history of \citet{robertson15}.  {\it Bottom panel:} Models that vary the final temperature after hydrogen reionization. We show two cases that have linear histories with $6<z_{\rm rei}<9$ and that vary the temperature after reionization between $T_{\rm rei} = 10,000~$K and $T_{\rm rei} =30,000~$K.  We also show a model with the ionization history from \citet[$6<z_{\rm rei}<13$ and $T_{\rm rei} =30,000~$K]{daloisio15}.  The points with error bars are the two highest redshift temperature estimates of~\citet{becker}. The red points have been corrected for pressure smoothing and the black points are uncorrected.}
\label{HI2}
\end{figure}

The curves in Figure~\ref{HI2} model the more realistic case of an extended hydrogen reionization process. To create these extended histories, we average a series of instantaneous reionization calculations at different times in different locations with reionization spanning $6<z<9$, $8<z<11$, and $6<z<13$. Indeed, averaging together $T(\Delta)$ from a series of instantaneous reionization calculations largely approximates how reionization is thought to have occurred;  ionization fronts sweep over the Universe, ionizing and heating up different regions at different times.  Afterwards, an ionized region cools via the well known physics described previously.  To generate the measured $T(\Delta)$, we then average the temperature of these different histories for a given global reionization history, an approach justified in Section~\ref{ss:comparing}. 

  The top panel of Figure~\ref{HI2} shows models with post-reionization temperatures of $20,000$ K and reionization durations of $6 < z_{\rm rei} < 8$, $6 < z_{\rm rei} < 9$, and $8 < z_{\rm rei} < 11$, where the \HII\ fraction evolves as a linear function of redshift with $x_{\rm HII}=0$ at $z=z_{\rm rei}^{\rm max}$ and $x_{\rm HII}=1$ at $z= z_{\rm rei}^{\rm min}$. This is a multi-phase model.  The instantaneous heating is associated with ionization, so the fraction that is heated follows the ionized fraction. We also include a reionization scenario from \citet{robertson15} that spans $6< z_{\rm rei} < 12$.  We find that a reionization scenario ending at $z=6$, with even a modest duration of $\Delta z=3$, is consistent with the measurements. However, these late-ending reionization models become inconsistent with the measurements when the duration is shorter than $\Delta z \approx 2$. The bottom panel shows models with different temperatures after reionization. The dashed curves both span $6<z_{\rm rei}<9$, but have $T_{\rm rei} = 10,000$ and $30,000$ K, respectively.  The solid curve shows a reionization scenario from \citet{daloisio15} that spans $6 < z_{\rm rei} < 13$, with $T_{\rm rei} = 30,000$ K. This model was found to generate the opacity fluctuations seen in the $z\sim 5.5$ Ly$\alpha$ forest \citep{becker15}. Though the resulting $T_0$ is dependent on the shape of the ionization history, in general, as long as reionization is extended, with $\Delta z \gtrsim 5$, $T_{\rm rei} = 30,000$ K is fully consistent with the \citet{becker} temperature measurements at $z= 4.8$ and $T_{\rm rei} = 20,000$ K is consistent for any history with $\Delta z \gtrsim 3$.   

\section{Temperature evolution during \HeII\ reionization}
\label{sec:general}

 Quasars are likely responsible for the reionization of helium, with the duration of this process shaped by their emissivity history.  Indeed, the known population of quasars produce roughly the correct number of \HeII\ ionizing photons to reionize the Universe at $z\sim3$ \citep{furlanetto09}, an epoch when observations of the \HeII\ Ly$\alpha$ forest strongly suggest that \HeII\ reionization was ending \citep{mcquinnGP, worseck11, shull10}.  In addition, investigations into other sources of \HeII\ ionizing photons have found that they are unlikely to be sufficient (\citealt{furlanetto09}, although see \citealt{miniati04}).  The combinations of quasars having rather hard spectra and the potentially long mean free path of \HeII\ ionizing photons ($\lambda_{\rm MFP, HeII}$), with
\begin{equation}
\lambda_{\rm MFP, HeII}\approx5\,\bar{x}^{-1}_{\rm HeII }\left(\frac{E_{\gamma}}{100\rm eV}\right)^3\left(\frac{1+z}{4}\right)^{-2} {\rm comoving~Mpc},
\label{eqn:mfp}
\end{equation}
where $E_\gamma$ is the photon energy, results in a different structure for the photoheating that occurs during \HeII\ reionization compared to hydrogen reionization \citep{mcquinn09}. 

We model the photoheating during \HeII\ reionization with a multi-zone model that approximates the heating process found in detailed simulations  \citep{mcquinn09}.  First, the lower energy photons with shorter $\lambda_{\rm MFP, HeII}$ are assumed to be consumed in producing a \HeIII\ bubble around quasars, instantaneous heating the ionized gas by, on average,
\begin{equation}
\Delta Q^{\rm inst}_X=n_{X}\left(\int_{h\nu_{X}}^{E_{\rm max}}dE\frac{J_E}{E}\right)^{-1}\int_{h\nu_{X}}^{E_{\rm max}}dE(E-h\nu_X)\frac{J_E}{E}, \label{eqn:Qinst}
\end{equation}
where $X$ is the relevant species, $J_E\propto E^{-\alpha}$ is the average specific intensity that ionized the gas, $h\nu_X$ is the ionization potential of the ion (54.4\,eV for \HeII), and $E_{\rm max}$ is the maximum energy photon that is typically absorbed in the process of making the \HeIII\ bubble (i.e. $\lambda_{\rm MFP}$ less than the bubble size).  Equation~(\ref{eqn:Qinst}) makes the approximation that all photons between $h\nu_X$ and $E_{\rm max}$ are absorbed within the bubble, an approximation motivated by how strongly $\lambda_{\rm MFP}$ depends on $E_{\gamma}$.  The gas temperature then increases by $\Delta T_{\rm inst}$ where $3/2 k_b n_{\rm tot} \Delta T_{\rm inst} = \Delta Q_{\rm inst} $ once the \HeIII\ ionization front passes.  For our fiducial value of $E_{\rm max}$, $\Delta T_{\rm inst}$ is approximately $\{8100, 8500, 8900\}$ K at $\alpha_{\rm QSO}=\{1.7, 1.5, 1.3\}$. Because these fronts take time to fill intergalactic space, different gas parcels experience this temperature increase at different times set by the duration of \HeII\ reionization. 
    
The $E_\gamma >E_{\rm max}$ photons stream through the Universe with mean free paths much greater than the bubble size.  While they do not contribute much to the ionization, the photoionizations from these hard photons heat the IGM in an approximately uniform manner with a rate of
\begin{equation}
\frac{dQ^{\rm hard}_{\rm HeII}}{dt}(z)=n_{\rm HeII}(z)\int_{E_{\rm max}}^{\infty}\frac{dE}{E}(E-E_{\rm HeII})J_E(z)\sigma_{\rm HeII}(E).
\label{eqn:hardbackground}
\end{equation}
Here the specific intensity of this hard background at redshift $z_0$ is
\begin{equation}
J_E(z_0)=\frac{c}{4\pi}\int_{z_0}^{\infty}dz\left|\frac{dt}{dz}\right|\frac{(1+z_0)^3}{(1+z)^3}\epsilon_E(z)e^{-\tau_{\rm HeII}(z_0,z,E)},
\label{eqn:JE}
\end{equation}
where we parameterize the specific emissivity as $\epsilon_E=AE^{-\alpha}$ [erg s$^{-1}$cm$^{-3}$eV$^{-1}$]. The normalization factor $A$ can be solved for by requiring that the total ionizing emissivity of ionizing photons balances the number of ionizations plus recombinations:
\begin{equation}
\int_{E_0}^{\infty}dE\epsilon_E= \bar n_{\rm He}\left(\frac{d\bar{x}_{\rm HeIII}}{dt}+C_{\rm HeIII}\alpha_B\bar{x}_{\rm HeIII} \bar n_e\right),
\label{eqn:normalization}
\end{equation}
where $\bar{x}_{\rm HeIII}$ is the instantaneous mean ionization fraction of \HeIII, and $C_{\rm HeIII}$ is a clumping factor enhancement in the recombination rate over a homogeneous Universe with $T=10^4$K.  Both $\bar{x}_{\rm HeIII}(z)$ and $C_{\rm HeIII}$ are specified in our model.  The specific intensity of this background depends on optical depth to redshift $z$ of photons emitted at $z_{\rm em}$:
\begin{align}
 & \tau_{\rm HeII}(z_{\rm em},z,E)  = \nonumber \\  &  \int_{z_0}^{z_{\rm em}}\frac{c \, dz}{H(z)(1+z)}\sigma_{\rm HeII}\left(E\frac{1+z}{1+z_0}\right)\bar{n}_{\rm HeII}(z).
\label{eqn:tau}
\end{align}
Regions that become ionized by a local quasar will have the photoheating from the hard photon background terminated, as there are no more \HeII\ ions to ionize.

Our implementation of \HeII\ reionization is based on applying these two photoheating processes to an ensemble of gas parcels.  Different gas parcels in the ensemble experience \HeII\ reionization at different times. When the first parcel becomes reionized, the remaining parcels begin to experience heating from the hard photon background. This is subsequently shut off in a parcel as soon as it becomes reionized and experiences instantaneous heating. Allowing individual regions to heat and subsequently cool separately provides a more realistic scenario than a one-zone model which implements uniform heating from a UV background (as done in \citealt{ewaldetal}).  In what follows, the two heating regimes are delineated by an upper limit to short mean free path photons of $E^{\rm inst}_{\rm max}=150~$eV.  We find that our results are negligibly changed by varying $E^{\rm inst}_{\rm max}$ over the plausible range of $100$ to $200\,$eV with a maximum temperature variation of $\sim1500$ K, corresponding to $\lambda_{\rm MFP} = 5-40~$comoving Mpc at $z=3$ (eqn.~\ref{eqn:mfp}).

Figure~\ref{Hires_Tarrs} shows the values of $T(\Delta)$ in a characteristic model at redshifts before and during \HeII\ reionization. The values of $T(\Delta)$ follow a tight power-law until \HeII\ is reionized in the first regions.  These regions are then heated by $\sim 8000$ K. As new regions become ionized, the first \HeIII\ bubbles cool, introducing spread in the temperature-density relation.  In addition, the hard background progressively builds up and heats the regions in which the \HeII\ has yet to be ionized.  The maximum dispersion in the temperature at fixed $\Delta$ in our models is found at $z\sim3$ when \HeII\ reionization terminates.  Afterwards the IGM cools and unshocked gas is driven to a single temperature-density relation \citep{1997MNRAS.292...27H}.  The distribution in temperatures in our models is similar to that found in radiative transfer simulations of \HeII\ reionization \citep{mcquinn09} and \citet{compostella13}.

\begin{figure}
\begin{center}
\hspace{-1.15cm}
\resizebox{9.0cm}{!}{\includegraphics{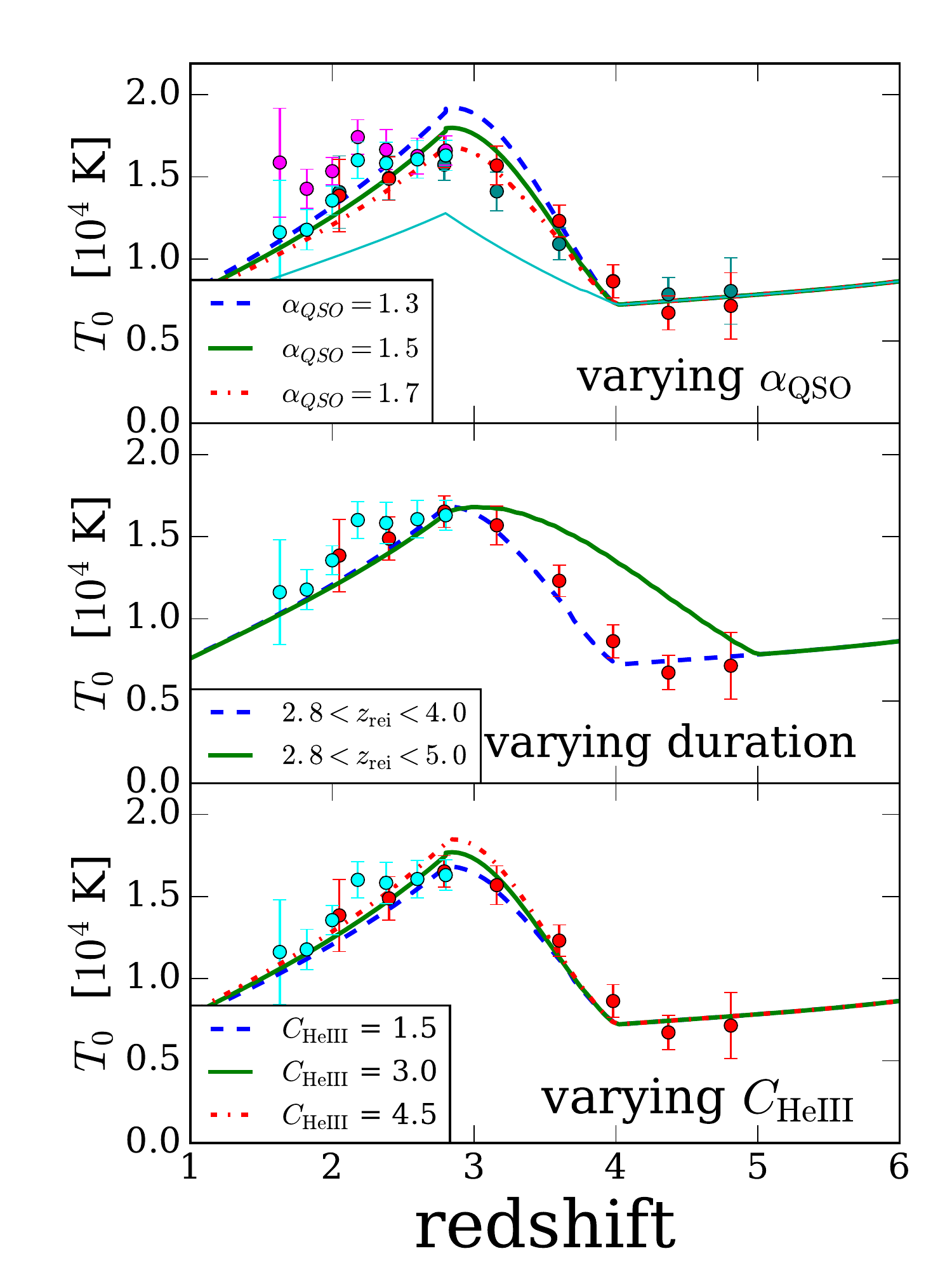}}\\
\end{center}
\caption{Average temperature of the IGM at the cosmic mean density for the case in which \HeII\ reionization has a \HeIII\ fraction that is linear over  $2.8<z<4.0$. Top panel: Models that vary the spectral index of quasars during \HeII\ reionization fixing $z_{\rm max, HeII}=4.0$ and $C_{\rm HeIII}=1.5$. The bottom thin, teal curve corresponds to a thermal history with the fiducial parameters and the hard background photons turned off. Middle panel: Models that vary the duration of \HeII\ reionization fixing $C_{\rm HeIII}=1.5$ and $\alpha_{\rm QSO}=1.7$. Bottom panel: The effect of varying the recombination rate, $C_{\rm HeIII}$, on the temperature during \HeII\ reionization fixing $\alpha_{\rm QSO}=1.7$ and $z_{\rm max, HeII}=4.0$. The points with error bars are the measurements of ~\citet{becker} and \citet{boera}. In the top panel, the red Becker points have a pressure smoothing corrections and the green Becker points have no correction. The cyan Boera points have an optical depth adjustment and the magenta points do not (see Appendix~\ref{ap:corrections}). All of the plotted models assume that \HI\ reionization instantaneously heats the IGM to $T=20,000~$K at $z = 9$, but we show in \S~\ref{sec:beforeHeIIreion} that our results depend weakly on this assumption.}
\label{HeII_1}
\end{figure}

In Figure~\ref{HeII_1} we calculate the temperature at mean density, $T_0$, in a simplistic model in which the global \HeIII\ fraction is a linear function of redshift. This means that the \HeIII\ fraction evolves as a linear function of redshift with $x_{\rm HeIII}=0$ at $z= z_{\rm rei, HeII}^{\rm max}$ and $x_{\rm HeIII}=1$ at $z= z_{\rm rei, HeII}^{\rm min}$. Unless otherwise noted, we assume fiducial parameters of $\alpha_{\rm QSO}=1.7$ for the spectral index of quasars' intrinsic emission, $\alpha_{\rm bk}=1.0$ for the spectral index of the ionizing background, $2.8<z_{\rm rei, HeII}<4$ for the duration of \HeII\ reionization, and $C_{\rm HeIII}=1.5$ for the \HeII\ clumping factor (which is proportional to the recombination rate).  The top panel of Figure~\ref{HeII_1} varies the spectral index of quasars between $\alpha_{\rm QSO}=1.3$ and $\alpha_{\rm QSO}=1.7$, the middle panel varies the duration of \HeII\ reionization such that reionization spans $2.8<z_{\rm rei}<4$ and $2.8<z_{\rm rei}<5$, and the bottom panel varies the clumping factor of \HeIII\ gas over $1.5 \leq C_{\rm HeIII} \leq 4.5$.  
The range of quasar spectral indexes is motivated by the range of constraints from the stacking analyses \citep{telfer02, shull12, lusso15}.  The range of \HeII\ reionization redshifts is motivated by the histories found in analytic calculations and simulations \citep{furlanetto09, mcquinn09}.  The range of clumping factors $C_{\rm HeIII}$ is motivated by the simulations in \citet{2012MNRAS.426.1349M}, in which intergalactic absorbers were exposed to a wide range of \HeII\ ionizing backgrounds.  In each panel, the observational measurements (points with error bars) have been extrapolated to $T_0$ using this fiducial model's $T-\Delta$ relation.  Most models are in striking (qualitative) agreement with the measurements.  Figure~\ref{HeII_1} shows that the thermal history of the IGM is mildly affected by the spectral index of the quasars, and it is insensitive to changes in the clumping factor for a physically motivated range of values.  However, the middle panels shows that the thermal history is very sensitive to the duration of \HeII\ reionization.  The measurements of \citet{becker} and \citet{boera} clearly favor the case with $2.8<z_{\rm rei, HeII}<4$ over the case with $2.8<z_{\rm rei, HeII}<5$.\footnote{It should be noted that our models' predicted $T_0$ are slightly lower than in the HeII reionization simulations of \citet{mcquinn09}.  The \citet{mcquinn09} models predicted that the peak in $T_0$ falls between 16,000 and 20,000K, with the exact value depending on the run specifications (whereas our fiducial model yields a peak of approximately 15,000K).  We attribute these differences to McQuinn et al.'s relatively hard choice of $\alpha_{\rm bk} =0$, their initialization temperature at z=6 being higher than in our models (where we more consistently model the high-redshift temperature evolution), and their calculations ignoring free-free cooling (which results in a ~1K difference).}

\begin{figure}
\begin{center}
\hspace{-1.15cm}
\resizebox{9.0cm}{!}{\includegraphics{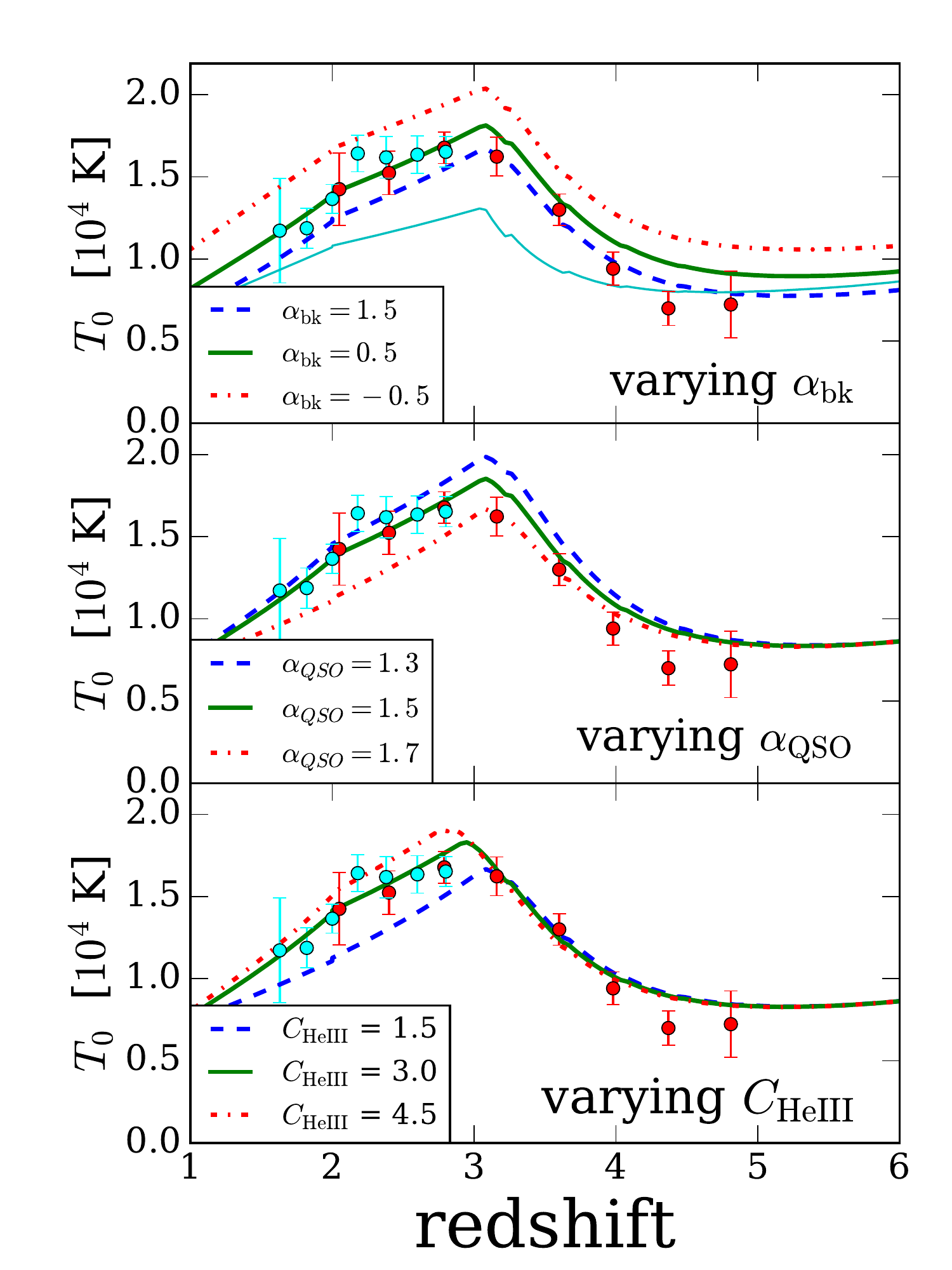}}\\
\end{center}
\caption{Average temperature of the IGM at the cosmic mean density for the case in which the \HeIII\ fraction is determined by the quasar emissivity as calculated from the quasar luminosity function of \citet{hopkins07}.  Top panel: Models that vary the background spectral index fixing $\alpha_{\rm QSO}=1.7$ and $C_{\rm HeIII}=1.5$. The bottom thin, teal curve corresponds to a thermal history with the fiducial parameters and the hard background photons turned off. Middle panel: Models that vary the spectral index of quasars fixing $\alpha_{\rm bk} = 1.0$ and $C_{\rm HeIII}=1.5$. Bottom panel: Models that vary the recombination rate through the clumping factor, fixing $\alpha_{\rm QSO}=1.7$ and $\alpha_{\rm bk} = 1.0$. The points with error bars are the measurements of ~\citet{becker} and \citet{boera}. The Becker points have a pressure smoothing corrections and the Boera points have an optical depth adjustment (See Appendix~\ref{ap:corrections}).  All of the plotted models assume that \HI\ reionization instantaneously heats the IGM to $T=20,000~$K at $z = 9$, but we show in \S~\ref{sec:beforeHeIIreion} that our results depend weakly on this assumption.}
\label{HeII_2}
\end{figure}

In Figure ~\ref{HeII_2} we adopt a reionization history that uses the fitting formula for the quasar emissivity given in \citet{haardtmadau}.  This formula is based on constraints on the \citet{hopkins07} luminosity function.  Unlike the previous models that assumed a linear-in-redshift \HeII\ reionization history, we now use this formula to self-consistently solve for the \HeII\ fraction also assuming a clumping factor to determine the recombination rate in \HeIII\ regions.  The top panel of Figure ~\ref{HeII_2} varies the spectral index of the \HI\ ionizing background between $\alpha_{\rm bk}= -0.5$ and $\alpha_{\rm bk}= 1.5$, the middle panel varies the intrinsic spectral index of quasars between $\alpha_{\rm QSO}=1.3$ and $\alpha_{\rm QSO}=1.7$, and the bottom panel varies the clumping factor from $C_{\rm HeIII}=1.5$ to $C_{\rm HeIII}=4.5$. (Increasing $C_{\rm HeIII}$ over this range delays the end of \HeII\ reionization by $\Delta z \approx 0.3$.) As before, we assume fiducial parameters of $\alpha_{\rm QSO}=1.7$, $\alpha_{\rm bk}=1.0$, and $C_{\rm HeIII}=1.5$, unless otherwise noted, and the observational measurements have been extrapolated to $T_0$ using this fiducial model's $T-\Delta$ relation.  These models with more empirically motivated \HeII\ reionization histories than the previous linear histories also show qualitative agreement with the measurements.  In detail, we find the thermal history in this models is most sensitive to $\alpha_{\rm bk}$:  The \citet{becker} and \citet{boera} measurements across $3 < z \lesssim 5$ favor the models with the softest spectrum of $\alpha_{\rm bk}\approx 1.5$.  On the other hand, the measurements at redshifts $z\lesssim3$ are somewhat above our fiducial thermal history (the green solid curve in all panels) and well above the $\alpha_{\rm bk} = 1.5$ dashed curve in the top panel.  A hardening of the background spectrum as quasars become increasingly dominant with decreasing redshift could explain some of this trend.  In addition, in the next section we address the possibility that this excess owes to heating mechanisms beyond the photoheating we have so far considered.

\section{Non-standard heating processes}
\label{sec:nonstandard}
Beyond the standard processes already discussed, some have speculated additional heating mechanisms may contribute to the temperature evolution of the IGM,  such as heating from cosmic rays\footnote{We note that cosmic rays may be more likely to pressurize rather than heat the IGM depending on their energy spectrum \citep{lacki13}.  Whether they heat the IGM depends on their energy which sets their loss timescale, with losses occurring most efficiently around $\sim1$MeV \citep{samui05}  In the case where they only pressurize the IGM, they would show up in the pressure-smoothing dependence of Ly$\alpha$ lines but not in their thermal broadening.  Ly$\alpha$ forest temperature measurements are sensitive to both effects, although more sensitive to the latter.  Thus, constraints on the non-thermal pressurization of the IGM from $\langle |\kappa| \rangle$ measurements would likely be much weaker.} \citep{samui05, lacki13}, from the intergalactic absorption of blazar TeV photons\footnote{The blazar TeV photons interact with radiation backgrounds, producing $e\pm$ pairs. The IGM heating is driven by streaming instabilities between the pair beam and IGM \citep{broderick12}.  However, see \citet{miniati13} and \citet{sironi14} for arguments why these instabilities may not occur, although we note that there is still an active debate (A. Broderick, private communication).} \citep{broderick12, chang12, puchwein12}, from broadband intergalactic dust absorption \citep{inoue08}, or from the byproducts of dark matter annihilations \citep{cirelli09}.  Here we constrain how much heating could owe to these processes. 

\begin{figure}
\begin{center}
\resizebox{9cm}{!}{\includegraphics{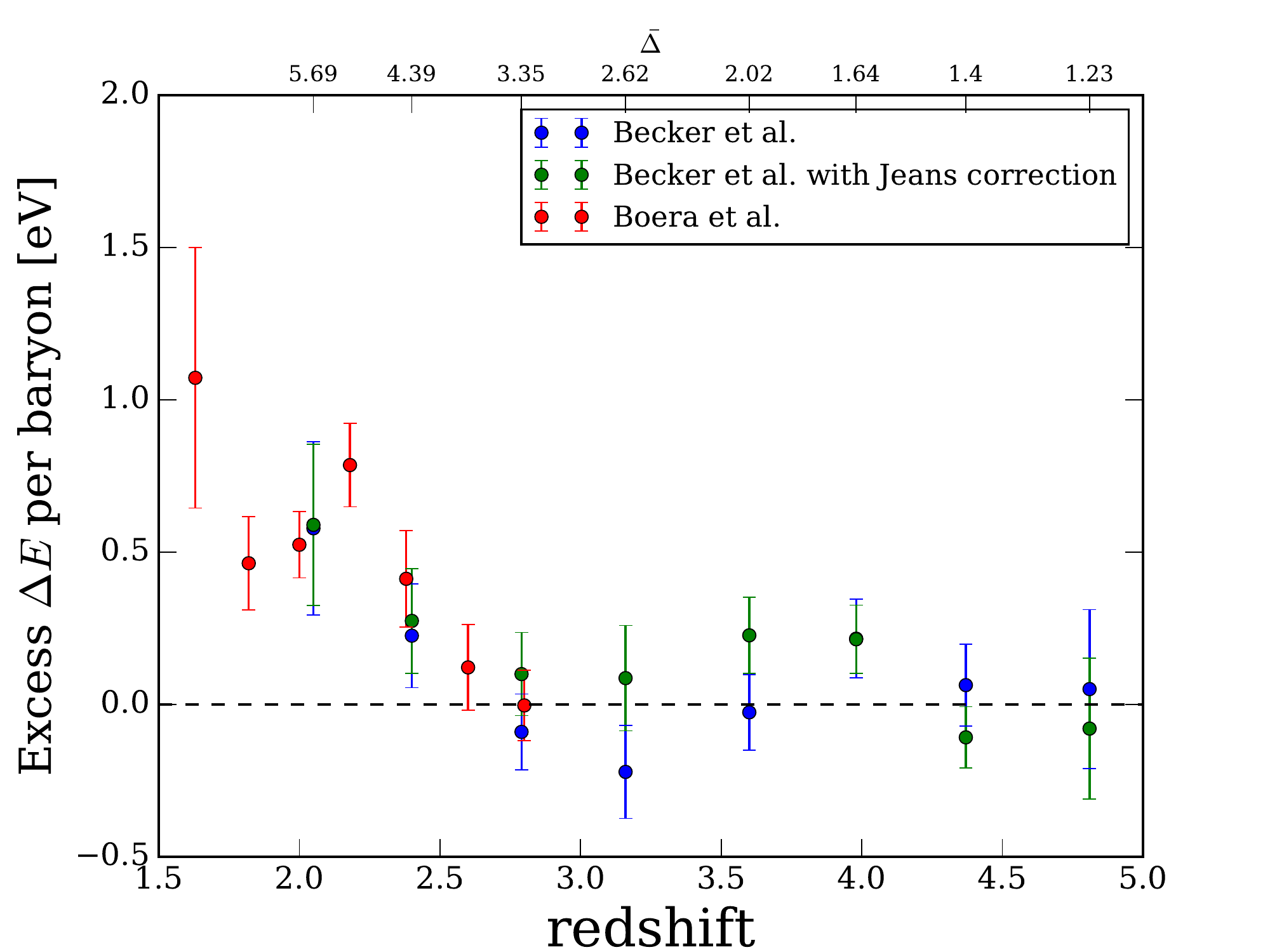}}\\
\end{center}
\caption{The excess energy per particle at $\Delta_{\star}$ between the measurements and our model with minimal photoheating.  This excess provides an upper bound on the amount of energy that can be injected by different non-standard heating models.}
\label{excess_eV}
\end{figure}

We first consider the excess heating that the temperature measurements allow over a model with the minimum possible amount of photoheating. In particular, to generate this minimum-photoheating model, we take our fiducial model ($T_0 = 20,000~$K and $z_{\rm rei}=9$) with quasar and background spectral indices on the softer side of potential values, $\alpha_{\rm QSO} = 1.7$ and $\alpha_{\rm bk} = 1.0$ (see discussion in \S~\ref{sec:beforeHeIIreion}), and with the expected recombination rate also on the low side with $C_{\rm HeIII}=1.5$.  We remind the reader that the data is mostly insensitive to the details of hydrogen reionization\footnote{The exception to this insensitivity is that the thermal history may impact the pressure smoothing at $z\sim 4-5$, hence the inferred temperatures.}, for which we assume an instantaneous reionization process with $T_{\rm rei} = 20,000~$K at $z=9$. Figure~\ref{excess_eV} shows the excess energy per baryon above this minimum-photoheating model at the $\Delta_{\star}$ of the measurements, which are displayed on the top axis. We find that an excess energy of $1~$eV per baryon between $z=2$ and $z =3$ can be accommodated by the \citet{becker} and \citet{boera} measurements.   The allowed excess heating is smaller at higher redshifts.  

This constraint does not account for cooling and adiabatic heating.  Accounting for cooling would allow the amount of non-standard heating to be somewhat higher, especially if the heat injection were extended in time.  In what follows, we model the heating as a function of time to account for cooling.

The proposed non-standard heating mechanisms would inject heat with different temporal trends and in a manner that differs in which gas densities are heated up more.   With regard to density, the intergalactic absorption of blazar TeV photons is the most straightforward, injecting energy in a volumetric manner \citep{chang12, puchwein12}, although with a small level of spatial fluctuations \citep{lamberts15}.  Cosmic ray heating \citep{samui05, lacki13} and heating due to broadband intergalactic dust absorption \citep{inoue08} are more likely to scale with density to some power as they are sourced by star forming regions within galaxies (and the cosmic rays may adjust to achieve equipartition with the gas pressure).  With regard to their temporal scaling, blazar heating likely traces the history of supermassive black hole accretion, and cosmic rays and dust are more likely to trace the evolution of the star formation rate density.  Conveniently, the quasar emissivity history and the evolution of the star formation rate density are largely similar, with the star formation rate density falling off more slowly at higher redshifts (where the IGM is less sensitive to additional heating anyway).  Thus, a reasonable parametrization for a generic excess heating process is that it traces the quasar emissivity history and that it scales with density as $\Delta^n$ for $n\geq 0$.  We consider $n=0$ and $n=1$ here.

\begin{figure}
\begin{center}
\hspace{-1.15cm}
\resizebox{9.5cm}{!}{\includegraphics{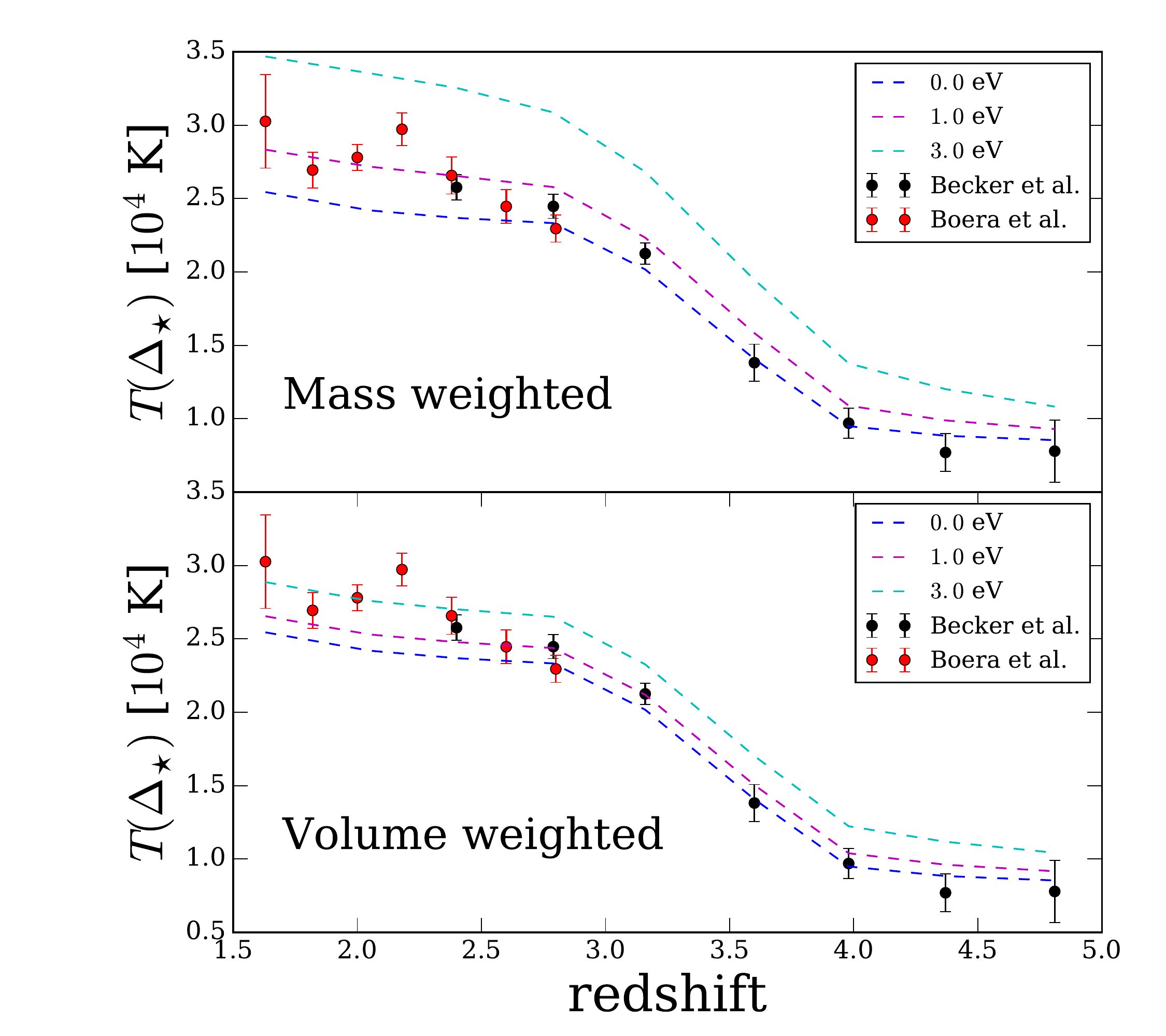}}\\
\end{center}
\caption{Temperature of the IGM at the densities corresponding to the \citet{becker} and \citet{boera} measurements with minimal photoheating and additional heat injection. This injection is normalized such that a total of either 0.0, 1.0 or 3.0 additional electron volts are added per particle at the mean density by $z=0$ over the minimal photoheating model in a manner that traces the quasar emissivity history, a parametrization motivated in the text. The top panel shows models in which the additional heat injection is mass weighted (the injected energy per particle is the same at all $\Delta$), and the bottom panel shows models where it is volume weighted (the injected energy per particle scales as $\Delta^{-1}$). }
\label{delta_quasar_tracing}
\end{figure}

\begin{figure}
\begin{center}
\resizebox{9.0cm}{!}{\includegraphics{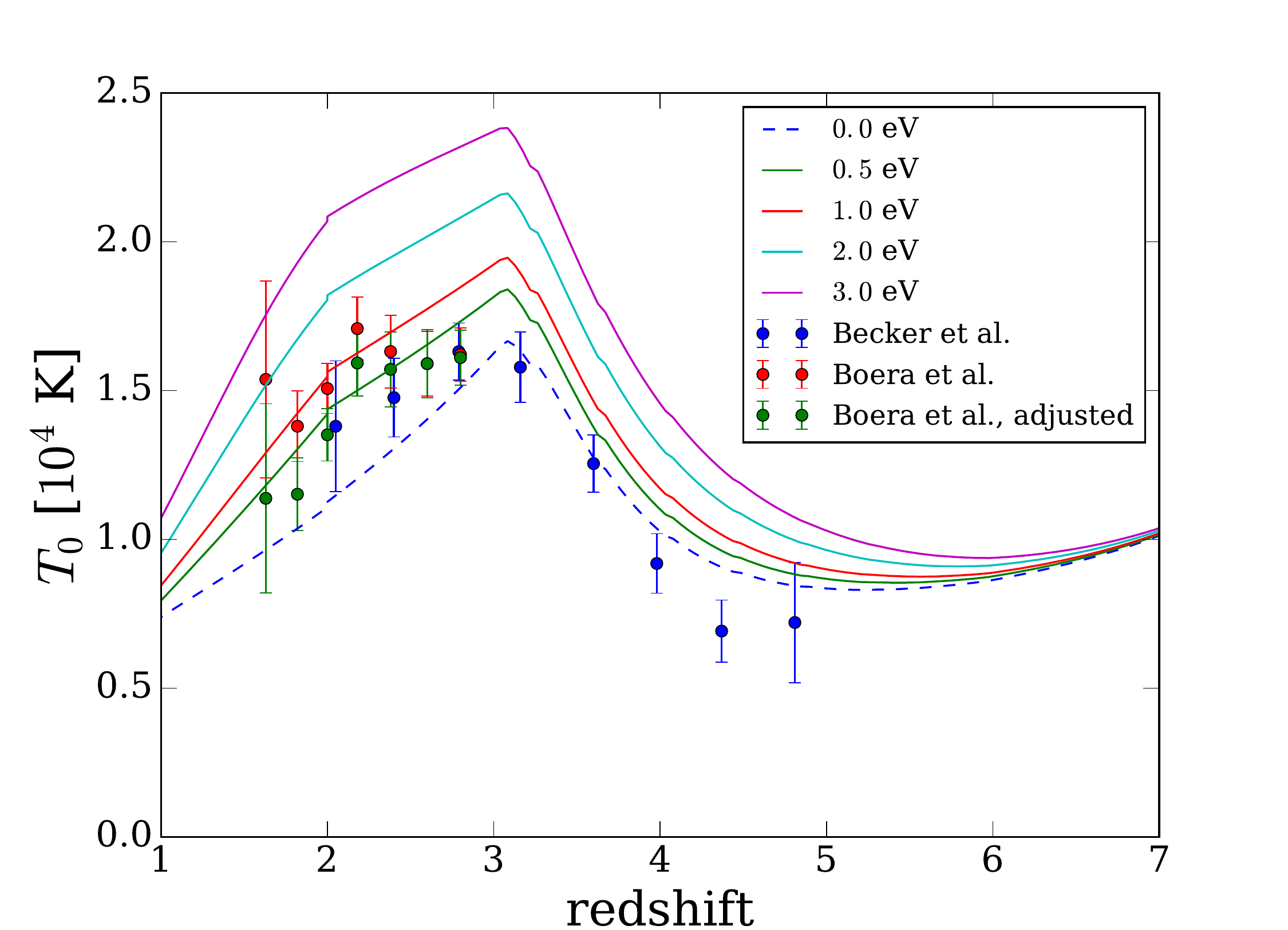}}\\
\end{center}
\caption{Average temperature of the IGM at the cosmic mean density in our minimal photoheating model and in models with additional heat injection. This injection is normalized such that 0.5, 1.0, 2.0, and 3.0 additional electron volts are added per particle by $z=0$ over the minimal photoheating model in a manner that traces the quasar emissivity history, a parametrization motivated in the text.}
\label{quasar_tracing}
\end{figure}

Figure~\ref{delta_quasar_tracing} adopts this parametrization, showing $T(\Delta_{\star})$ for different amounts of total heat injected into the IGM that are normalized over the quasar emissivity history such that 1.0 and 3.0 additional electron volts are added per particle at $\Delta=0$, with $\{33\%,62\%,90\%\}$ of the heating occurring by $z=\{4, 3, 2\}$.  The top panel injects this heating in a mass-weighted manner (scaling with $\Delta$) and the bottom a volumetric manner (no $\Delta$ dependence so the energy per particle scales as $\Delta^{-1}$).  We use the same functional form for the quasar emissivity as in Figure~\ref{HeII_2}, and again we use the minimal-photoheating parameters $\alpha_{\rm QSO} = 1.7$, $\alpha_{\rm bk} = 1.0$, and $C_{\rm HeIII}=1.5$. In addition, Figure~\ref{quasar_tracing} shows instead $T_0$ for different amounts of total heat injected into the IGM at the cosmic mean density rather than at $\Delta_{\star}$. The heat input is normalized such that 0.5, 1.0, 2.0, and 3.0 additional electron volts are added per particle over the entire quasar emissivity history. To extrapolate to $T_0$ from $T(\Delta_{\star})$, we use the $T-\Delta$ relation from the model with no additional heating input.  

Previously published models for blazar heating \citep{chang12, puchwein12}, cosmic ray heating \citep{samui05}, dust heating \citep{inoue08}, and dark matter annihilation heating \citep{cirelli09} have forecasted heat injections at $\Delta_{\star}$ of more than $1\;$eV per particle by the lowest considered redshifts.  (Blazar heating may be the least constrained by our analysis because it is volumetric, which only in the most extreme models of \citealt{puchwein12} result in $1\;$eV per particle at $\Delta_{\star}$ by $z= 2$.)  Thus, our limits place constraints on the parameter space of these models. However, there is not a lower bound on the heating predicted for any of these mechanisms and so our constraints do not rule out these processes contributing at the sub-eV level.

\section{Conclusions}

The standard picture for the thermal history of the IGM is that it is shaped by two major heating events -- the reionization of \HI\ and of \HeII.   We developed a model for the thermal history in this picture and compared it against recent observations to constrain these and other heating processes.

Regarding \HI\ reionization, we found that, because the temperature after reionization quickly cools to an asymptotic value, even the highest redshift temperature measurements (at $z=4.8$) are consistent with most possible \HI\ reionization scenarios.  However, we showed that the temperature measurements do rule out extreme reionization scenarios in which 1) the reionization process has a short duration with $\Delta z \lesssim 3$ and ends at $z\sim6$ and 2) a passing ionization front heats the gas to temperatures on the hotter side of the expected range, $T_{\rm rei} \approx 30,000$ K. On the other hand, we found that an extended reionization spanning $z=6-9+$ with $T_{\rm rei} =20,000$ K -- the most likely value -- is not ruled out by the high-redshift temperature measurements.

  In addition, we found that our models of \HeII\ reionization that follow the cannonical quasar emissivity history (e.g. \citealt{hopkins07}) are consistent with the temperature measurements.  These models result in the bulk of \HeII\ reionization occurring over $z=3-4$.  The measurements are not consistent with \HeII\ reionization scenarios where \HeII\ reionization started $\Delta z>0.5$ earlier than $z=4$.  For example, we found a scenario with a linear ionization history between $z=3-5$ produced temperatures at $z>4$ that are well in excess of measurements.  At a lower confidence level, our models are most consistent with the $z=3.5-4.8$ temperature constraints if the \HI-ionizing background is on the softer side of estimates ($\alpha_{\rm bkgd}\geq1$).  By lower redshifts, temperature measurements favor the ionizing background to harden or, instead, for there to be additional heating processes. % (which may just be structure formation shocks, which we have not modeled but should become increasingly significant).

Finally, we placed upper bounds on additional heating mechanisms such as blazar heating, heating from cosmic rays, broadband intergalactic dust absorption, and other mechanisms. At $z> 3.5$, we found that the \citet{becker} measurements allow little room for additional mechanisms ($\ll 1\,$eV per baryon).  On the other hand, towards lower redshifts we found an increasing excess energy per baryon beyond our minimum photoheating model, reaching a maximum excess of $\approx1~$eV per baryon at $\Delta_*$ by $z=2$.  

Previous analytic studies of the IGM temperature have been based on a one-zone approach in which reionization is modeled to be a homogeneous process.  These one-zone models have been used to constrain the reionization process when compared with IGM temperature measurements \citep{haehnelt98}.  The one-zone models of \citet{theuns02c} and \citet{2003ApJ...596....9H} were used to show that the temperature measurements prefer either reionization to occur after $z\sim10$ or a $z\sim3-4$ reionization of \HeII.  More recently, \citet{raskutti12} used the redshift $z\sim 6$ temperature measurements from \citet{bolton12} to constrain reionization in the biased regions around quasars to have occurred between $z=6$ and $z=12$.  This redshift range is consistent with what we determine for the general IGM.  In addition, \citet{ewaldetal} modeled both hydrogen reionization and \HeII\ reionization. \citet{ewaldetal} used the photoheating rates from the \citet{haardtmadau} ionizing background model, supplemented with a non-equilibrium photoheating implementation for during \HeII\ reionization.  The \citet{ewaldetal} approach has the advantage that they could run cosmological simulations with these heating rates and, hence, directly compare the curvature measurements of \citet{becker} and \citet{boera} to those from mocks constructed using the simulations.  However, their approach has the disadvantage that one-zone models do not capture how intergalactic gas is heated and cooled during and after reionization processes.  We also expanded upon their work by using a range of models (e.g. instead of assuming the rigid \citealt{haardtmadau} form) to explore how uncertain inputs (such as the timing of reionization processes) affect the thermal history.  Perhaps because of the more physical implementation, our models perform better at describing the \citet{becker} than those in \citet{ewaldetal} as well as earlier one-zone models.

Future improvements in the understanding of the high-redshift IGM could come from several fronts. Firstly, temperature measurements at higher redshifts would better constrain hydrogen reionization processes \citep{2014ApJ...788..175L, daloisio15}. Additionally, the pressure smoothing of the IGM may be measurable with close pairs of quasar sightlines \citep{2013ApJ...775...81R}.  Finally, the \HeII\ Ly$\alpha$ forest may be able to better constrain the duration of \HeII\ reionization \citep{worseck14}.  The agreement between some recent temperatures measurements and the models presented here suggests that we are zeroing in on a concordance model for the IGM thermal history.\\

\section*{Acknowledgments}
We would like to thank the anonymous referee for their comments. We would like to thank E.~Boera for useful conversations and we would also like to thank G.~Becker for useful comments on the manuscript. This work is supported by the National Science foundation through grants AST-1312724 and AST-1514734 and by NASA through a grant from the Space Telescope Institute, HST-AR-13903.00.
\bibliography{temp_hist1}

\appendix

\section{Corrections to the measurements}
\label{ap:corrections}

This appendix addresses in more detail two correction factors discussed and used in the main text. The first is the way we account for the pressure smoothing in our temperature histories.  The second is the adjustment to the densities of the \citet{boera} measurements.

\subsection{Accounting for pressure smoothing}
\label{ap:jeans}

Because pressure smoothing extends the spatial distribution of hotter gas, with the smoothing depending on how far sound waves can travel, the line widths of the Ly$\alpha$ forest spectra are sensitive to the temperature history and not just the instantaneous temperature that thermal broadening probes.  This makes the measured ``temperature'' sensitive to the thermal history assumed. In \citet{becker} (Section 4.5), two models that mimic \HeII\ reionization (T15fast and T15slow) were used to demonstrate the bias of pressure smoothing on the inferred simulation. In particular, the bias between the curvature-method recovered temperatures (which was calibrated on artificial temperature histories) from the T15fast and T15slow runs and the actual T15fast and T15slow simulation temperatures demonstrate the degree to which pressure smoothing skews the measurements for these particular histories.  Because our model with a linear reionization over $2.8<z<4.0$ is similar in shape to the T15slow model, we use the T15slow pressure smoothing corrections to also correct our measurements as described in \S~3.2. 

\begin{figure}
\begin{center}
\resizebox{9.3cm}{!}{\includegraphics{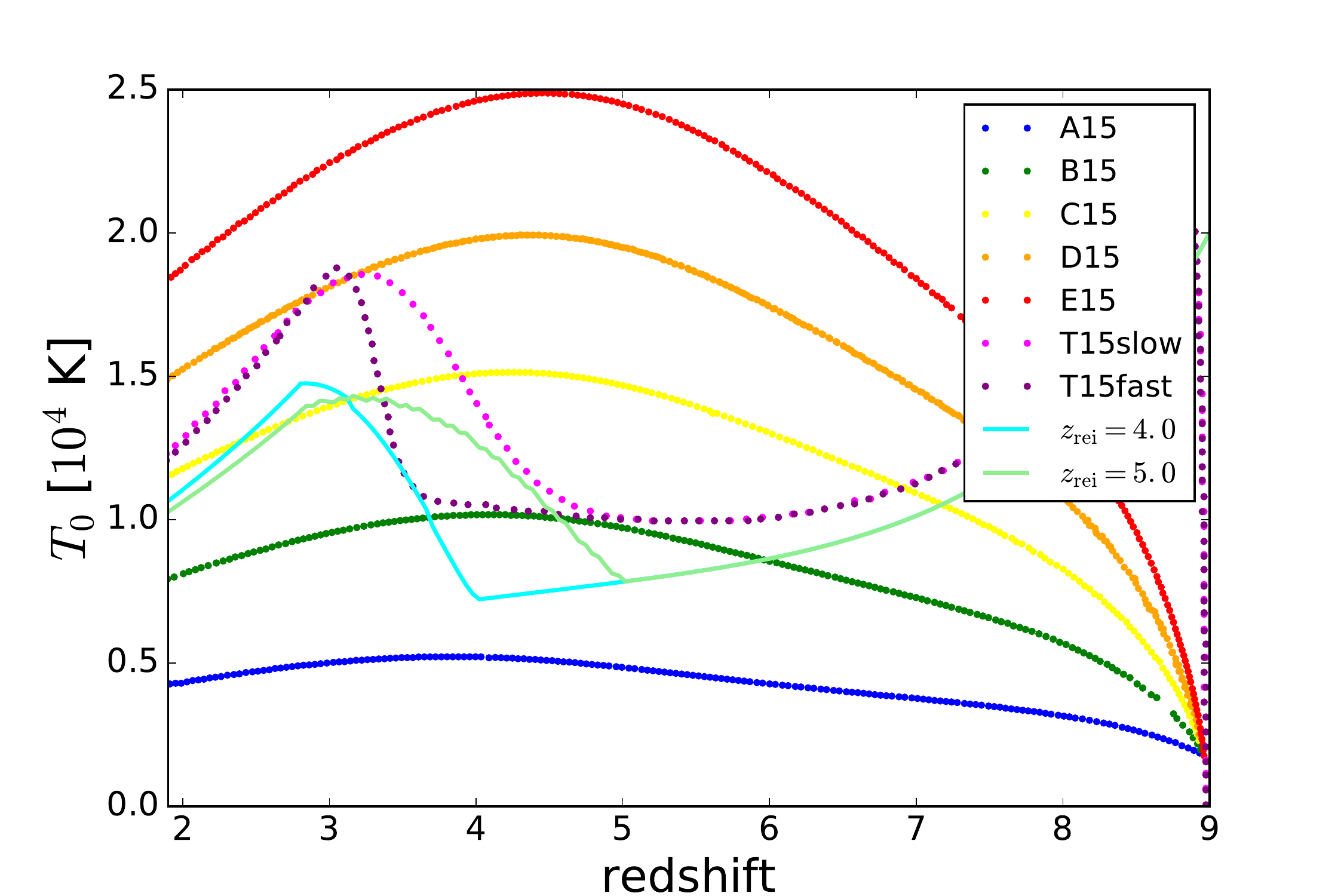}}\\
\end{center}
\caption{Models A15-E15, T15slow, and T15fast from \citet{becker} with the two curves from the middle panel of Figure~\ref{HeII_1}. Our two curves have linear \HeII\ reionizations scenarios and $C_{\rm HeIII}=1.5$, $\alpha_{\rm bk} = 1.0$, $\alpha_{\rm QSO}$, with  $z_{\rm max, HeII}=4.0$ and $z_{\rm max, HeII}=5.0$ respectively.}
\label{beckercompare}
\end{figure}

Figure~\ref{beckercompare} shows the ionization histories \citet{becker} used to calibrate their curvature measurements, their Models A15-E15, as well as their T15fast and T15slow runs.  We also show our models with a linear ionization history over  $2.8<z<4.0$ (characteristic of most of the \HeII\ reionizaiton calculations), and with $\alpha_{\rm QSO}=1.7$, $C=1.5$, and $\alpha_{\rm bkgd}=1.0$.  The duration of this model is most similar to T15slow, motivating our correction method (see \S~3.2).  We also show our most extended \HeII\ reionization history with the same parameters except $2.8<z<5.0$, which has a longer duration that T15slow.  The corrections for this case will be smaller than used in the text. In Section~\ref{ss:comparing}, we provide quantitative assessments of these corrections.

\subsection{Optical depth adjustments to the Boera measurements}
\label{ap:fluxcor}

In the main body, in addition to the original \citet{boera} measurements, we show a set of measurements with corresponding adjusted $\Delta_*$ that have been recalibrated to use the same evolution of mean optical depth as in \citet{becker} (Boera, private communication), rather than the internal mean optical depth estimated in \citet{boera}. This adjustment affects the extrapolation of the \citet{boera} measurement to $T_0$, and it makes the \citet{becker} and \citet{boera} $T_0$ more consistent at overlapping redshifts.  The mean optical depth in \citet{becker} is estimated using a larger data set and agreed well with previous measurements \citep{faucher08}, motivating this correction.

\end{document}